\documentclass[authoryear,round,sort&compress,preprint,5p]{elsarticle}

\usepackage{graphicx}
\usepackage{amssymb}

\usepackage{subfigure}
\usepackage{booktabs}
\usepackage{multirow}
\usepackage{multicol}
\usepackage{tabularx}
\usepackage{rotating}
\usepackage{amsmath}
\usepackage{algorithm}
\usepackage{algorithmic}

\usepackage[pageanchor=false]{hyperref}
\usepackage{url}
\usepackage{xcolor}
\usepackage[]{pdfcomment}
\usepackage{xspace}
\usepackage{array}
\usepackage{balance}
\usepackage{float}

\usepackage{listings}
\usepackage{fancybox}
\newcommand{\mybox}[1]{
	\vspace{0.15cm}
    \begin{flushleft}
	\ovalbox{
        \begin{minipage}{0.45\textwidth}
            \vspace{0.2cm}
            #1
            \vspace{0.2cm}
        \end{minipage}
    }
    \end{flushleft}
}

\newtheorem{mydef}{Definition}

\journal{Journal of Systems and Software}

\begin{document}

\begin{frontmatter}

\title{From API to NLI: A New Interface for Library Reuse}

\author[1,2]{Qi Shen}
\author[1,2]{Shijun Wu}
\author[1,2]{Yanzhen Zou\corref{mycorrespondingauthor}}
\cortext[mycorrespondingauthor]{Corresponding author}
\ead{zouyz@pku.edu.cn}
\author[3]{Zixiao Zhu}
\author[1,2]{Bing Xie}

\address[1]{Key Laboratory of High Confidence Software Technologies, Ministry of Education, Beijing, 100871, China}
\address[2]{School of Electronics Engineering and Computer Science, Peking University, Beijing, 100871, China}
\address[3]{IBM Research, Beijing, 100191, China}

\begin{abstract}
Developers frequently reuse APIs from existing libraries to implement certain functionality.
However, learning APIs is difficult due to their large scale and complexity.
In this paper, we design an abstract framework \textsc{NLI2Code} to ease the reuse process.
Under the framework, users can reuse library functionalities with a high-level, automatically-generated NLI (Natural Language Interface) instead of the detailed API elements.
The framework consists of three components: a \emph{functional feature extractor} to summarize the frequently-used library functions in natural language form, a \emph{code pattern miner} to give a code template for each functional feature, and a \emph{synthesizer} to complete code patterns into well-typed snippets.
From the perspective of a user, a reuse task under \textsc{NLI2Code} starts from choosing a functional feature and our framework will guide the user to synthesize the desired solution.
We instantiated the framework as a tool to reuse Java libraries.
The evaluation shows our tool can generate a high-quality natural language interface and save half of the coding time for newcomers to solve real-world programming tasks.
\end{abstract}


\end{frontmatter}

\section{Introduction}
\label{intro}

To implement certain functionality, developers often reuse existing libraries with the corresponding APIs.
Yet discovering the correct subset of the APIs is a major obstacle for the API users \citep{api-hard2}.
The obstacle not only comes from the large size of APIs, furthermore, a real-world programming task usually requires the cooperation of multiple APIs, and each API invocation should follow strict specifications.
For example, for a simple functionality like ``\textit{set color for an Excel cell}", the desired API usage sequence using \texttt{apache-poi} is as follows:
\begin{equation}\nonumber
    \begin{split}
    &Workbook.createCellStyle();\\
    &CellStyle.setFillBackgroundColor(short);\\
    &CellStyle.setFillForegroundColor(short);\\
    &CellStyle.setFillPattern(FillPatternType);\\
    &Cell.setCellStyle(CellStyle);
    \end{split}
\end{equation}

\begin{figure}[!htb]
    \centering
    \subfigure[User types or selects the desired functional feature]{
    \begin{minipage}{8cm}
    \centering
    \includegraphics[width=\textwidth]{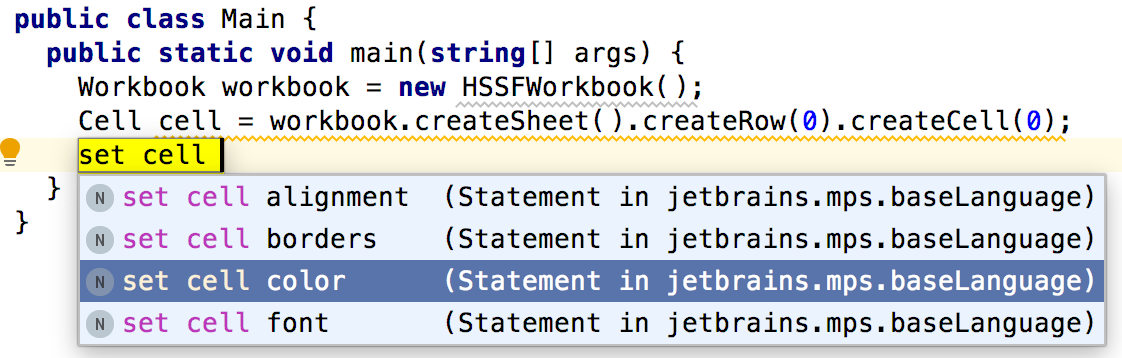}
    \vspace{0.1cm}
    \end{minipage}
    }
    \centering
    \subfigure[The code pattern exposes three parameters for user]{
    \begin{minipage}{8cm}
    \centering
    \includegraphics[width=\textwidth]{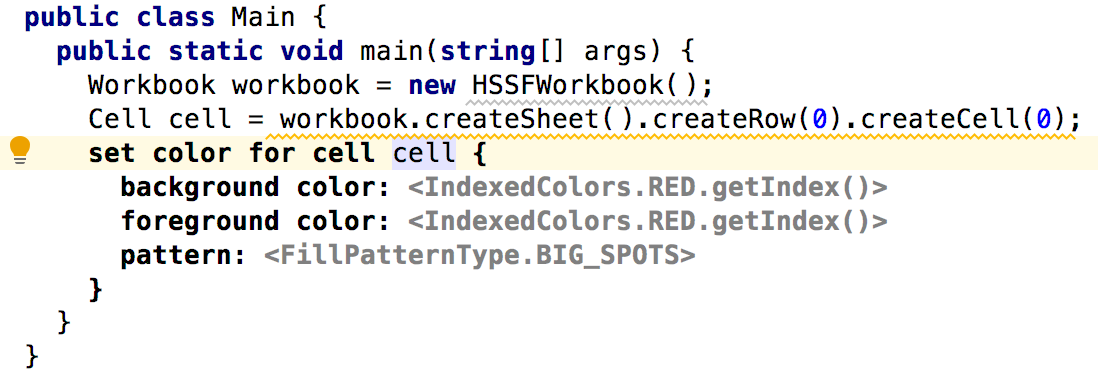}
    \vspace{0.1cm}
    \end{minipage}
    }
    \centering
    \subfigure[The synthesized code snippet for the functional feature]{
    \begin{minipage}{8cm}
    \vspace{0.2cm}
    \includegraphics[width=\textwidth]{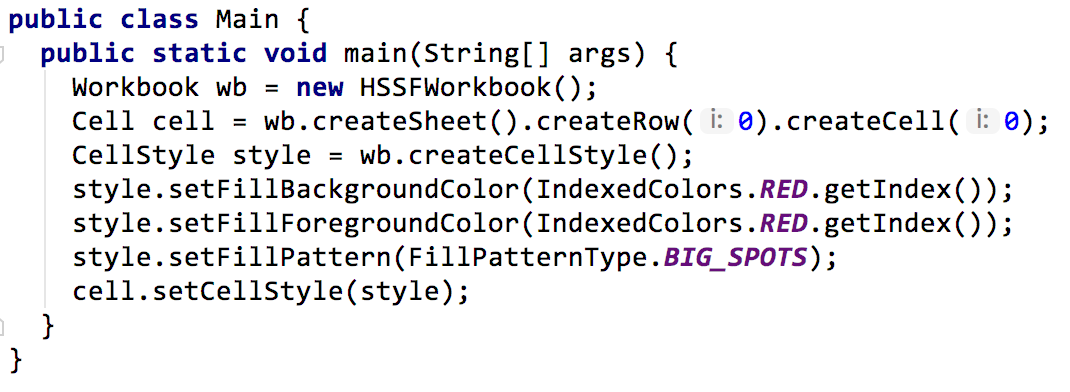}
    \end{minipage}
    }
    \caption{Application of NLI for reusing \texttt{apache-poi}}
    \label{mps}
\end{figure}

To address the issue, we promoted the concept of NLI (Natural Language Interface) for library reuse \citep{DBLP:conf/icsr/ShenXZZW19}.
With NLI, users reuse library functionalities with high-level natural language descriptions instead of directly manipulating the detailed APIs.
Figure \ref{mps} summarizes the key steps of how Alice, a Java programmer, reuses the library \texttt{apache-poi} with NLI.
As Figure \ref{mps}.(a) shows, Alice starts from selecting the desired functionality from a list of natural language descriptions, which is \emph{set cell color} in this case.
After the selection, the functionality is mapped to its corresponding implementation, which is a built-in code template in NLI.
As Figure \ref{mps}.(b) shows, Alice needs to provide three parameters (\textit{i.e.} specific background color, foreground color and fill pattern) to fill the template.
Each parameter to provide is annotated with an example expression in grey font, which is recommended by a synthesizer.
In fact, there are more than three missing parameters in the code template, the synthesizer has automatically created trivial ones from the current context (\textit{e.g.} creating a \textit{Workbook} object with the constructor).
After Alice fills the parameters, a well-typed code snippet is synthesized and inserted into the editor (as Figure \ref{mps}.(c) shows), which perfectly solves Alice's task.

For the library reuse problem, we highlight the benefits of NLI from two aspects.
The first benefit is the query composition.
If the developer is not familiar with the library, it could be difficult to compose a high-quality query which accurately describes the desired functionality.
For example, Table \ref{qa} displays a post\footnote{https://stackoverflow.com/questions/53052931} from Stack Overflow.
The post title mistakenly mentioned the concept ``\textit{background color}'', while the accepted answer shows the user actually desired ``\textit{foreground color}''.
In NLI, we summarize the library functionalities into functional features.
We conjecture that, compared to composing free-form queries, the mechanism of selecting functional features is easier and can make users more confident about the results.
The second benefit is code quality.
An illustrative code example can help developers quickly understand how to implement certain functionality.
However, many online code examples are only intended to express the main idea of a solution instead of being reused as-it.
Previous studies \citep{DBLP:conf/icsm/TreudeR17, api-misuse} show that online code examples are often not self-explanatory and may have quality problems such as incorrect order of API calls.
As Table \ref{qa} shows, the code snippet in the accepted answer contains only one API, which is not a complete solution for the task.
In NLI, we mine code patterns by exploring more usage examples of the API.
Our hypothesis is that unveiling how APIs are used in similar program contexts could improve the code quality.

\begin{table}[t]
    \centering
    \caption{An example post from Stack Overflow}
    \label{qa}
    \begin{tabular}{p{0.46\textwidth}}
    \hline
    \textbf{Title: Apache-POI : How to set background color of a cell when creating spreadsheet?}\\
    \hline
    \textbf{Question: } In Apache POI 4.0, I want to set an Excel cell background color.
    But all I get are black cells. I've tried many things, but result is always the same.
    How can I set the background color of an Excel cell in Apache POI 4.0 ?\\
    \hline
    \textbf{Answer: }Try to use below code for background style:\\
    \small{setFillForegroundColor(IndexedColors.YELLOW.getIndex());}\\
    \hline
    \end{tabular}
\end{table}

To construct and use NLI, we designed an abstract framework \textsc{NLI2Code}, which consists of three components: a functional feature extractor, a code pattern miner, and a synthesizer.
\emph{Functional features} are natural language descriptions of the library functionalities.
In this paper, we instantiated the extractor by mining Stack Overflow since a previous survey shows it is the first option for most developers to search for programming solutions \citep{uclsurvey}.
In the second component, we try to match each functional feature with a \emph{code pattern}, which is a code template mined from multiple implementations of the feature.
As code patterns usually lack customized information such as local parameters, a \emph{synthesizer} is supposed to complete them into compilable snippets.
The missing parameters could be synthesized from the current programming context or provided by the user.
Finally, \textsc{NLI2Code} combines the three components and generates well-typed code snippets for users.

Around the central concept \textsc{NLI}, the main contributions of this paper are:
\begin{itemize}
\item an algorithm to extract verb phrases describing library functionalities from Stack Overflow.
\item an approach to mine code patterns, with a self-designed intermediate representation for Java to eliminate coding style differences.
\item an instantiation of \textsc{NLI2Code} to reuse Java libraries, with evaluation on real-world tasks to prove the feasibility of the framework.
\end{itemize}

The remainder of the paper is organized as follows.
Section \ref{framework} demonstrates the abstract framework \textsc{NLI2Code}.
Sections \ref{feature}, \ref{pattern} and \ref{synthesizer} explain our implementation of the framework, which is available from our online artifacts \footnote{https://github.com/nli2code/jss-artifact}.
In Section \ref{evaluation}, we conduct several experiments to check the accuracy of our algorithms and apply a controlled experiment to explain how \textsc{NLI2Code} works in real-world development.
Section \ref{related} introduces the related work.
Section \ref{conclusion} briefly summarizes this paper.
\section{Framework}
\label{framework}

\begin{figure}
\includegraphics[width=0.48\textwidth]{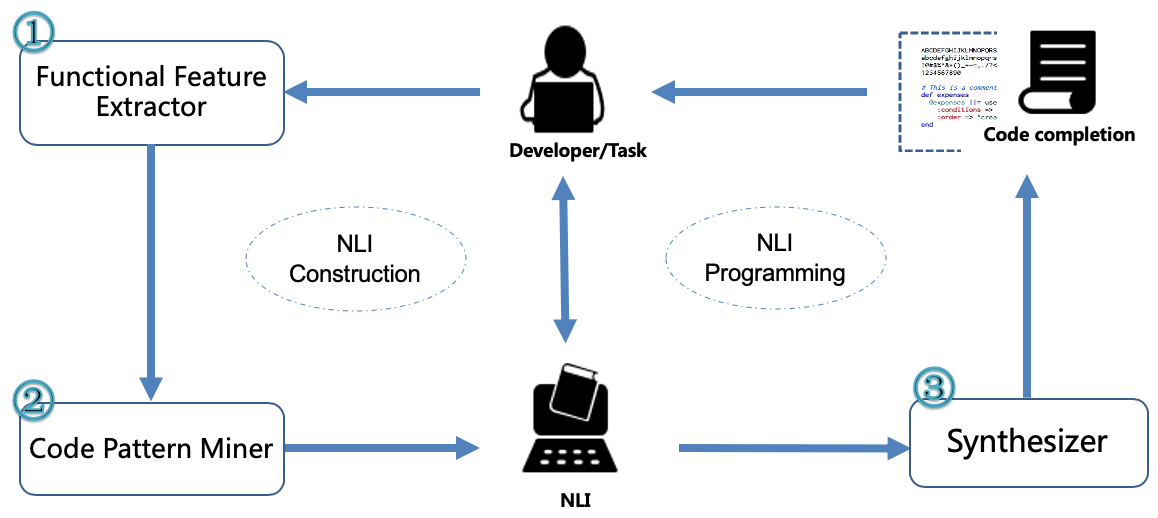}
\caption{NLI2Code framework}
\label{fig-framework}
\end{figure}
As Figure \ref{fig-framework} shows, the \textsc{NLI2Code} framework consists of three components.
In the offline part, we construct a natural language interface as pairs of functional features and code patterns.
In the online part, the user solves tasks by selecting functional features and our synthesizer completes the corresponding code patterns into well-typed code snippets.
In the rest of this section, we will discuss the three components separately.

\subsection{Functional Feature Extractor}
\begin{figure}
\includegraphics[width=0.45\textwidth]{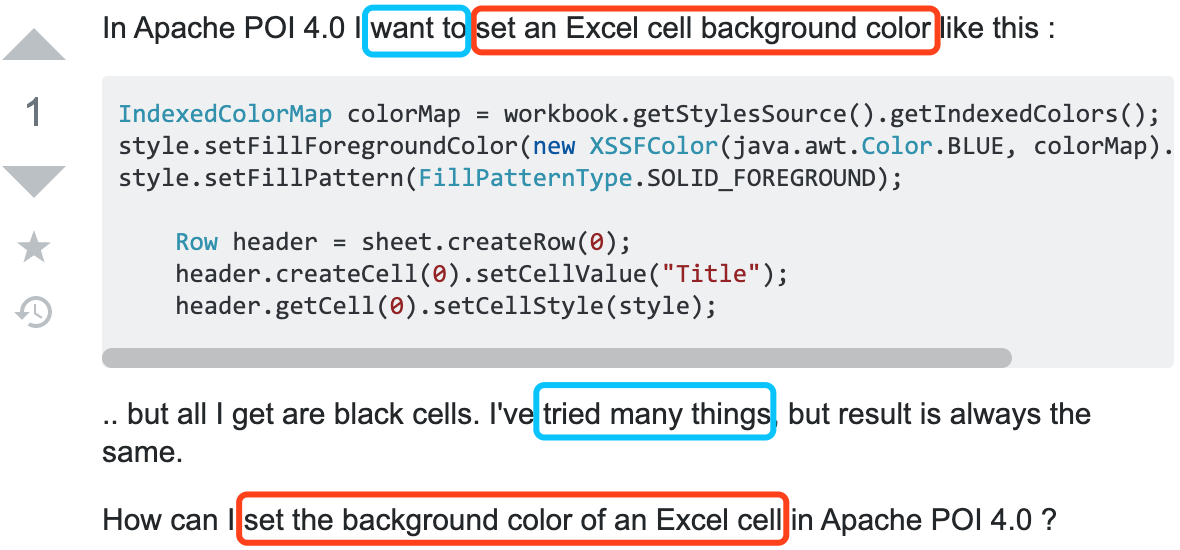}
\caption{Functional features in Stack Overflow}
\label{fig-so}
\end{figure}

Extracting functional features is the first step in our framework.
A functional feature is a brief description of certain library functionality in verb phrase form.
Nowadays, libraries typically provide multiple platforms for developers and users to communicate, such as mailing lists, issue tracker system, and posts from online forums like Stack Overflow.
These communication records are the natural corpus to extract functional features because they contain rich information about how libraries are used.
In our framework, all verb phrases from the discussions are considered as candidate functional features.
We issue two challenges to get usable functional features:
\begin{itemize}
\item Noises. As Figure \ref{fig-so} shows, phrases like \emph{want to} and \emph{try many things} are unrelated to library functionalities and have little semantic information. Such phrases should be pruned off. 
\item Diversity. Functionalities could be expressed in different ways. \textit{e.g.} \emph{set an Excel cell color} and \emph{set the color of an Excel cell} in Figure \ref{fig-so}.
Furthermore, users could use different words which makes the phrases lexically different. \textit{e.g.} \emph{change the cell color}.
Such phrases with the same semantic information need to be clustered and normalized.
Otherwise, the generated natural language interface will be verbose and repetitive.
\end{itemize}

In this work, we applied a filtering pipeline to remove noise phrases, considering stop words, the structure, and the context of the phrases.
To cluster similar phrases, we designed a normal form to extract the core action and objects in verb phrases.
After normalization, phrases with the same content or merely lexically different are merged.
Here we define two important properties for the extracted functional features:
\begin{itemize}
\item Accurate: Each functional feature should clearly correspond to certain library functionality.
\item Complete: The set of all functional features should cover the library functionalities as much as possible.
\end{itemize}

\subsection{Code Pattern Miner}
Functional features organize library functionalities in a list of verb phrases.
Although many posts from the user forums naturally provide code examples, these examples usually cannot be reused as-it.
In fact, most code examples are only intended to describe the main idea of a solution to the original question, which could be difficult to be understood by others \citep{DBLP:conf/icsm/TreudeR17}.
Furthermore, a recent analysis shows that online code examples usually have quality problems such as missing control constructs and incorrect order of API calls \citep{api-misuse}.
Another analysis on 914,974 Java code snippets from Stack Overflow shows that only 3.89\% of them are parsable \citep{compilable}.

A practical way to improve the quality of code examples is to detect similar API usage in a larger codebase.
A code pattern is a code template describing that in a certain usage scenario, some API elements are frequently called together.
Compared to a single code example in the original post, a code pattern exploits the commonalities among similar programs, which reduces the risk of unknown consequences.
Moreover, code patterns naturally hint users which part of the code to modify because they leave variations among the programs as unfilled parts.
Common variations include hard-coded strings and magic numbers.
A common procedure for code pattern mining is as follows:
\begin{itemize}
\item construct a code corpus
\item abstract code into a certain data structure (\textit{e.g.}, call sequence, abstract syntax tree, data flow graph)
\item apply the corresponding frequent pattern mining algorithm on the corpus and transform the frequent items back to code
\end{itemize}

\subsection{Synthesizer}
Code patterns are incomplete because they usually miss local information.
Existing IDEs (Integrated Development Environment) usually provide a simple code completion feature.
However, such completion typically only considers one step of computation, which means that the recommendation result is a single variable or method.
In fact, a missing parameter may require a method chain to get the desired result.
Furthermore, each method in the chain may require new parameters to synthesize.
These efforts suggest a general direction for the synthesizer in \textsc{NLI2Code}:
given a programming context $\Gamma$ and the desired type $\tau$, synthesize the entire type-correct expression with type $\tau$ from the context.
Formally, find expression $e$ such that $\Gamma\vdash e: \tau$.

We conclude two solutions for the synthesizer.
The first one is called the type-directed search \citep{PLDI12:completion, insynth}, which enumerates all possible expressions with the desired type.
Since the searching space is usually large, heuristic functions are often used to guide the search process.
The second solution recommends expressions according to the statistical analysis of a large code corpus.
To synthesize the desired expression, users can benefit if they are recommended with expressions frequently used under a similar context. 

To understand the potential and feasibility of our framework, we instantiated it as a tool \textsc{NLI4j} to reuse Java libraries.
In the following three sections, we will separately introduce our implementation of the three components.
\section{Extracting Functional Features}
\label{feature}
\begin{figure*}[!htb]
\centering
\includegraphics[width=\textwidth]{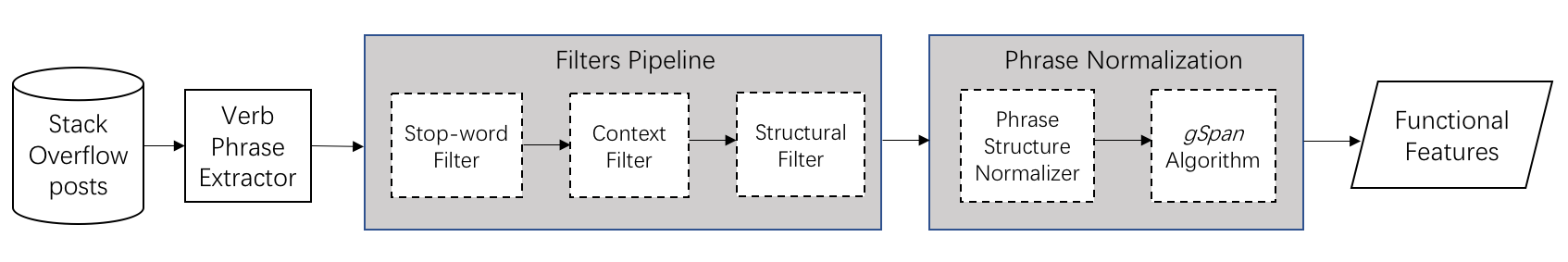}
\caption{The process of extracting functional features}
\label{fig-task}
\end{figure*}
As the first component of \textsc{NLI2Code}, we need a list of functional features to summarize frequently-used library functionalities.
Recall that a functional feature is defined as a brief description of certain functionality in verb phrase form.
Given a library, our extractor takes Stack Overflow threads as input and outputs the functional feature list.

Figure \ref{fig-task} shows the workflow of our approach to extract functional features.
We firstly extract verb phrases from Stack Overflow threads by leveraging the syntax parsing techniques.
Then, a set of heuristic rules is used to filter out unlikely phrases.
Considering that the same functionality can be expressed in different ways, we propose a normalized functional feature representation grammar to ensure the correct clustering of phrases.
At last, a frequent subgraph mining algorithm gSpan \citep{gSpan} is applied to mine functional features from the clustered phrases.

\subsection{Candidate Functional Features}

Our data source is the Q\&A threads from Stack Overflow containing the specific tags, such as \textit{``apache-poi''} for the POI project.
According to the definition, we extract all verb phrases as candidate functional features.
Similar to the state-of-the-art works, we use Stanford NLP toolkits \citep{stanford-nlp} to extract verb phrases from the raw data.
A big problem for applying NLP tools to software documentation is that there are many code-like terms, which are error-prone in the POS (Part-of-Speech) tagging and might cause failure in the syntax tree parsing.
Thus, we replace the code-like terms with special placeholders to ensure the correct POS tagging.
For reproducibility, we briefly explain how we recognize code-like terms here.
Stack Overflow threads usually label code fragments with consecutive $<pre><code>$ tags or $<code>$ tag for the inline code elements.
For those code-like terms that are not annotated with HTML tags, we employ a set of regular expressions, which is provided by \cite{task1} to identify them from the natural language content.

After the preprocessing, we split the natural language text into sentences and feed each sentence to the Stanford NLP toolkit.
The toolkit returns a tree-structured parsing result.
Figure \ref{fig-syntaxtree} displays the parsing tree of a long sentence, which contains seven verb phrases (subtrees tagged with VP).
All the verb phrases from the parsing tree are extracted and form the initial candidate functional features, which will be filtered by a filtering pipeline.
The pipeline filters out phrases not related to library functions, for example, in the sentence from Figure \ref{fig-syntaxtree}, only one phrase out of seven subtrees is valuable, which is the seventh phrase \emph{set up the print area for the excel file}. 

\subsection{Filtering Pipeline}
The filtering pipeline consists of three rule-based phrase filters.
If a phrase matches the rule of a filter, one piece of evidence will be added to the phrase.
A piece of evidence might be counted as one vote up or one vote down or veto to accept the phrase.
Then we collect all the evidence added to a phrase and count the vote.
Intuitionally, we remove a phrase when the upvotes are less than the downvotes.

The first filter is based on a handcrafting stop word list.
We downvote three types of verb phrases because they are not likely to appear in a meaningful functional feature: 
\begin{itemize}
\item Special grammatical ingredients such as auxiliary verbs (\textit{e.g.}, be, do, have), modal verbs and pronouns usually do not have actual meanings.
\item Q\&A special words. The sentences from Stack Overflow often contain trivial words for describing the questioners' requirements (\textit{e.g.}, ask, try, need).
\item Programming special terms. Some programming terms, keywords in programs, or development special words are usually not part of valid functional features. (\textit{e.g.}, extend, return and stack trace).
\end{itemize}

\begin{figure}[htb]
    \centering
    \includegraphics[width=0.43\textwidth]{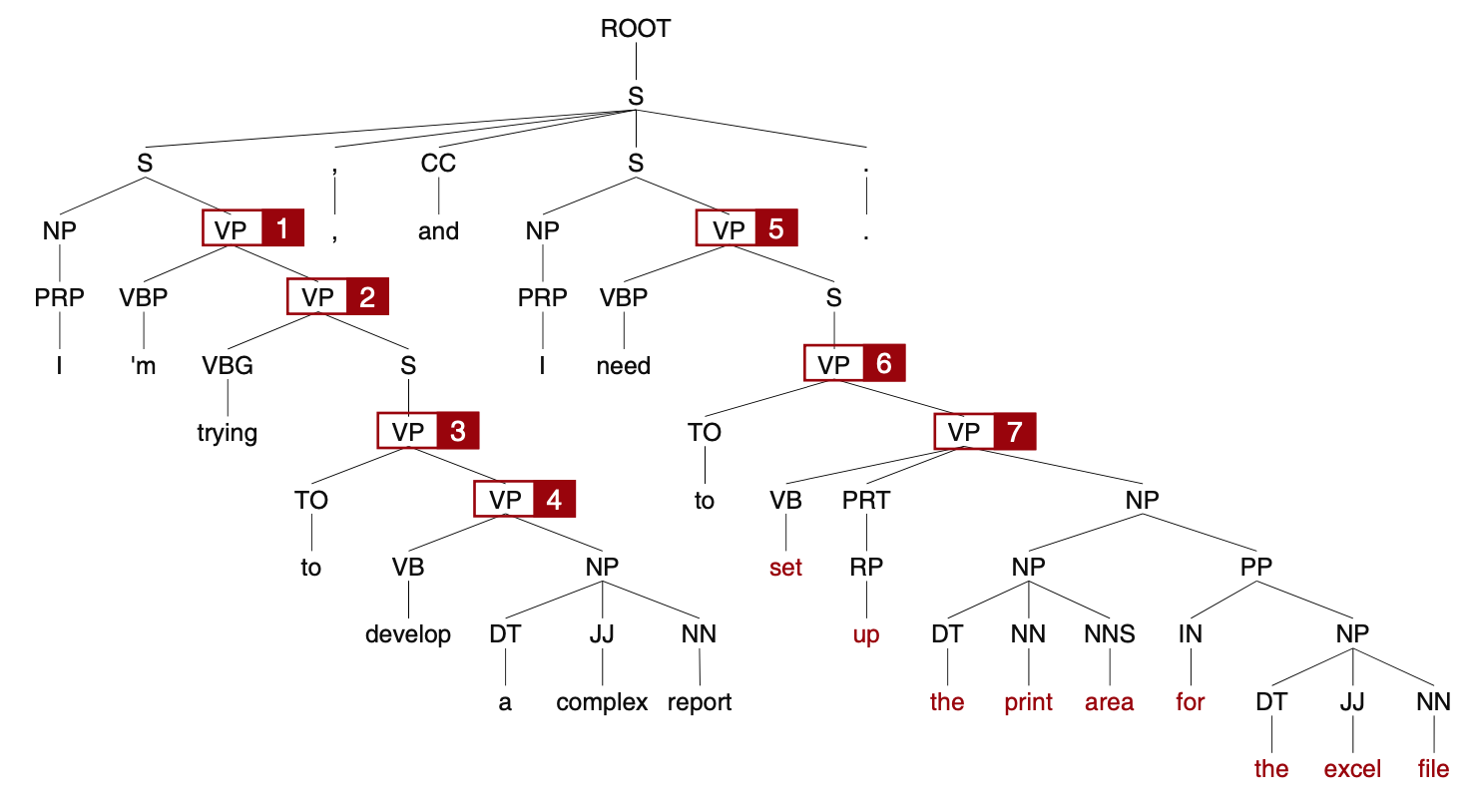}
    \caption{The parsing tree of a long sentence. The seventh verb phrase is a functional feature and the others need to be filtered.}
    \label{fig-syntaxtree}
\end{figure}

The second filter judges the phrases based on information from the context.
Though the phrases containing Q\&A special expressions are considered invalid, the phrases following some special Q\&A expressions are very likely to refer to the library functionalities.
For example, in Figure \ref{fig-syntaxtree}, the 5th verb phrase \emph{"need to ..."} should be filtered, but the 7th verb phrase \emph{"set up the print areas for the excel file"} following the Q\&A phrase \emph{"need to"} is a functional feature.
For each phrase, we analyze its preceding content in the same sentence.
If we find a match with Q\&A special expressions before the phrase, we upvote the phrase.

The third filter is based on the structure of the phrase in the syntax tree.
We use syntactic structure characteristics to filter out invalid verb phrases.
For example, the 3rd and the 6th phrases in Figure \ref{fig-syntaxtree} do not contain any verbs as direct children and will be filtered out with the structural filter.
Besides, there are usually some complex sub-clauses in the verb phrase.
We hope to keep our generated features as concise as possible, therefore we remove the sub-clauses.
Another important purpose of filtering parse tree structures is to get the candidate phrases ready for the later normalization.
The structural filter ensures the phrase candidates are compatible with the normal form.

\subsection{Phrase Normalization}
To cluster verb phrases with similar meaning, we define the normal form of feature phrases as Table \ref{bnf} shows.
The symbol ``[]" denotes that a component is optional.
Generally speaking, a functional feature consists of at least an \emph{Action} and an \emph{Object}, which could be modified by a \emph{Condition} (usually a prepositional phrase).

\begin{table}[htbp]
\caption{The normal form of feature phrases}
\label{bnf}
\centering
\begin{tabular}{l l l}
\hline
Feature & ::= & Action Object [Condition]\\
Action & ::= & verb [particle]\\
Object & ::= & dt adj noun\\
Condition & ::= & prep [verb] Object\\
\hline
\end{tabular}
\end{table}

Our pilot study summarizes the common parsing tree types that are compatible with our normal form.
To put this straight, Table \ref{transformrule} lists 6 types and their transformation rules to the normal form.
Case \#1 is the most common case, denotes the verb phrases that consist of a verb and a noun phrase.
Particles for the intransitive verbs are presented in case \#2.
Case \#3 is another popular case that contains a verb, noun phrase (NP), and prepositional phrase (PP).
Case \#4 denotes the verb phrases that do not contain a direct noun phrase.
Case \#5 is for the noun phrases that consist of a word chain headed by a noun.
Case \#6 is a typical prepositional phrase.

\begin{table*}[htbp]
    \centering
    \scriptsize
    \caption{Transformation rules for common types of verb phrases\label{transformrule}}
    \begin{tabular}{l l l l}
    \hline
    \bfseries ID & \bfseries Grammar pattern & \bfseries Phrase example & \bfseries Transformation rule to normal form\\
    \hline
    \multirow{2}{*}{1}&\multirow{2}{*}{VP:=VB NP}&get the cached formula value & \multirow{2}{*}{VB:=verb; NP:=Object}\\
    & & VP(VB)(NP(DT)(JJ)(NN)(NN)) &\\
    \hline
    \multirow{2}{*}{2}&\multirow{2}{*}{VP:=VB PRT NP}&set up the print areas & VB:=verb;PRT:=particle;\\
    & & VP(VB)(PRT(RP))(NP(DT)(NN)(NN)) & NP:=Object\\
    \hline
    \multirow{2}{*}{3}&\multirow{2}{*}{VP:=VB NP PP}&delete documents from lucene index & VB:=verb;NP:=Object;\\
    & & VP(VB)(NP(NN))(PP(IN)(NP(NN)(NN))) & PP:=Condition\\
    \hline
    \multirow{2}{*}{4}&\multirow{2}{*}{VP:=VB PP}&iterate through the terms in a document & VB:=verb;IN(in PP):=particle;\\
    & & VP(VB)(PP(IN)(NP(NP(DT)(NN))(PP(IN)(NP(DT)(NN))))) & NP(in PP):=Object\\
    \hline
    5&NP:=word NN&Case \#1-\#3& Map word to dt adj noun in Object\\
    \hline
    \multirow{2}{*}{6}&\multirow{2}{*}{PP:=IN NP}& \multirow{2}{*}{Case \#3-\#4} & IN ::= prep (in Condition);\\
    & & & NP ::= Object (in Condition)\\
    \hline
    \end{tabular}
\end{table*}

\begin{figure}[htb]
\centering
\includegraphics[width=0.43\textwidth]{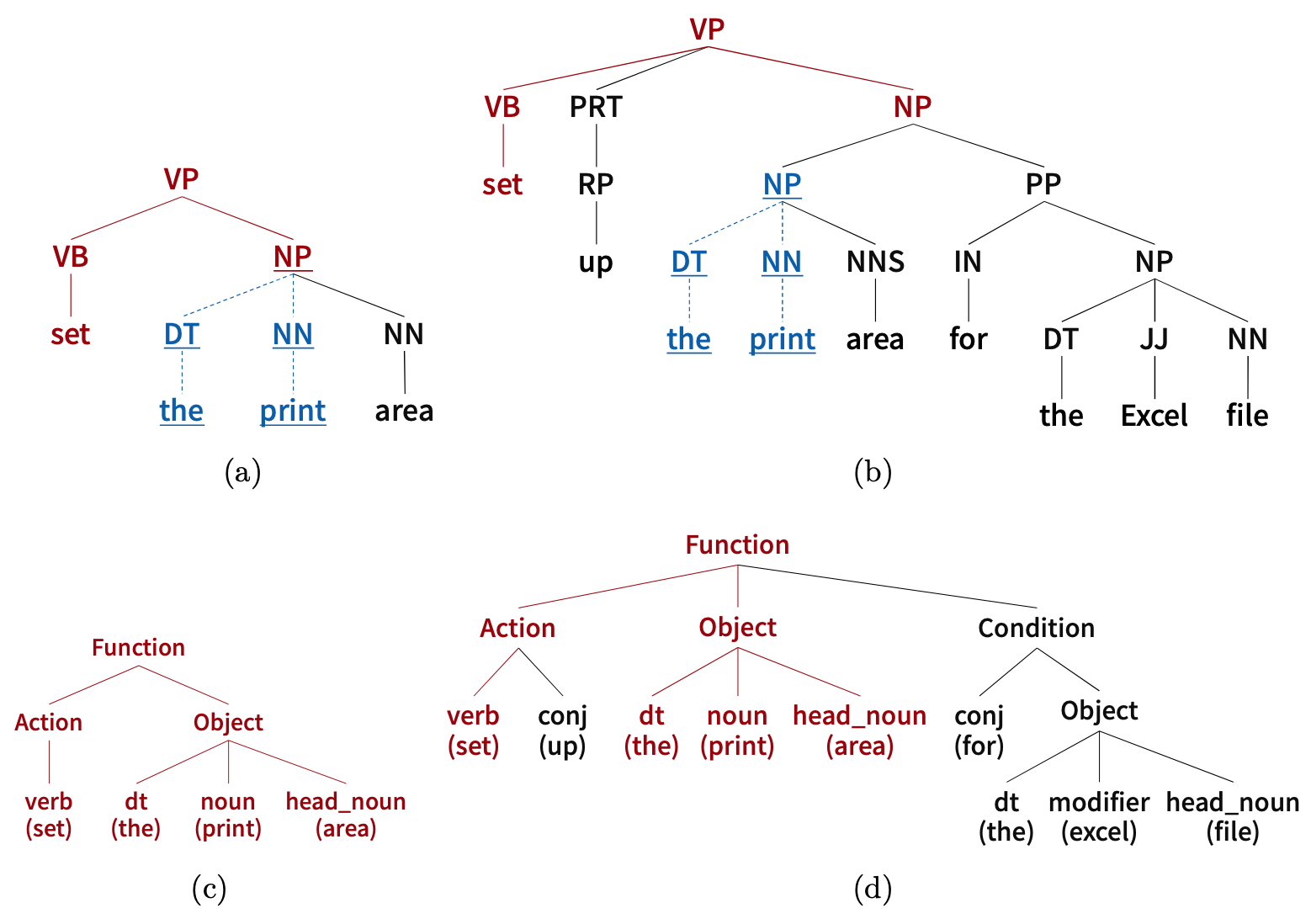}
\caption{Comparison between parse tree and normalized tree in mining frequent subtrees}
\label{cluster}
\end{figure}

After normalization, we rebuild the tree representation for the phrase and apply \emph{gSpan} algorithm to mine frequent subgraphs as our final functional features.
Furthermore, we merge two phrases if they share the same objects and their action words are synonyms judged by WordNet \footnote{We use APIs from nltk.corpus.wordnet}.
Figure \ref{cluster} explains why normalization is necessary.
Figure \ref{cluster}.(a) is the parse tree of the verb phrase \emph{set the print area} and Figure \ref{cluster}.(b) depicts another candidate phrase \emph{set up the print areas for the excel file}.
The original parsing trees contain many detailed grammatical ingredients, which prevent us from mining valuable common subgraphs.
The two largest common subgraphs between tree (a) and tree (b) are (VP (VB set) (NP)) (in red color and bold font) and (NP (DT the) (NN print)) (in blue color and underscored), which are meaningless.
In contrast, Figure \ref{cluster}.(c) and Figure \ref{cluster}.(d) are rebuilt from our normalized phrases, which omit unnecessary details like POS tags and unify the structures of the top layers.
Their common parts (in red and bold font) show us a reasonable result.

\section{Mining Code Patterns}
\label{pattern}
After getting the list of functional features, we map the features to their implementation.
Although Stack Overflow often provides direct code snippets along with the descriptions, such examples are usually incomplete (\textit{i.e.} only mentioned the key APIs instead of the complete solution) and may have quality problems such as incorrect order of API calls \citep{api-misuse}.
To augment these code examples, our main idea is to unveil what has been done in more similar programs.
To be specific, we first map each functional feature to a related API and construct a code corpus containing usage examples of the API.
Then, we abstract each code example in the corpus into a data flow graph and apply existing frequent subgraph mining algorithm to mine the patterns.
Finally, we transform the mined patterns (\textit{i.e.}, frequent subgraphs) back to the text-form code.

\subsection{Code Corpus Construction}
We first match each functional feature with a related API and then construct a code corpus by searching usage examples of the API.
The rationale behind this design decision is that although code snippets on Stack Overflow suffer the quality problems, they often mentioned the correct API to use.
Such APIs could be a starting point to find the complete solution.

Given a functional feature, we view all the code elements mentioned in the same Stack Overflow thread (\textit{i.e.}, contents inside the $<code>$ tag) as candidate APIs to match.
The metric we use to select the related API is based on the lexical similarity.
First, we split the API names according to the camel-case rule and stem the splitted tokens.
For each API, we calculate the number of overlapped tokens between its name and the functional feature. (\textit{e.g.}, the number is 2 for the API \textit{``setFillForegroundColor''} and the feature \textit{``set cell color''} since the overlapped tokens are \textit{``set''} and \textit{``color''}).
The API with the most overlapped tokens is selected as the matched one and we break the tie by counting the occurrence number of certain API in the thread.

After selecting a related API, we further extract usage examples of the API from client repositories downloaded from Github in advance.
If a source code file from the repositories contains the desired API, we add it to the corpus.

\subsection{Code Abstraction}

\begin{figure}[htb]
\lstset{language=Java}
\begin{lstlisting}
// snippet 1
style.setFillForegroundColor(short);
style.setFillPattern(SOLID_FOREGROUND);
// snippet 2
style.setFillPattern(SOLID_FOREGROUND);
style.setFillForegroundColor(short);
\end{lstlisting}
\caption{Example snippets where the sequence model fails}
\label{poi-order}
\end{figure}
    
\begin{figure}[htb]
    \centering
    \includegraphics[width=0.45\textwidth]{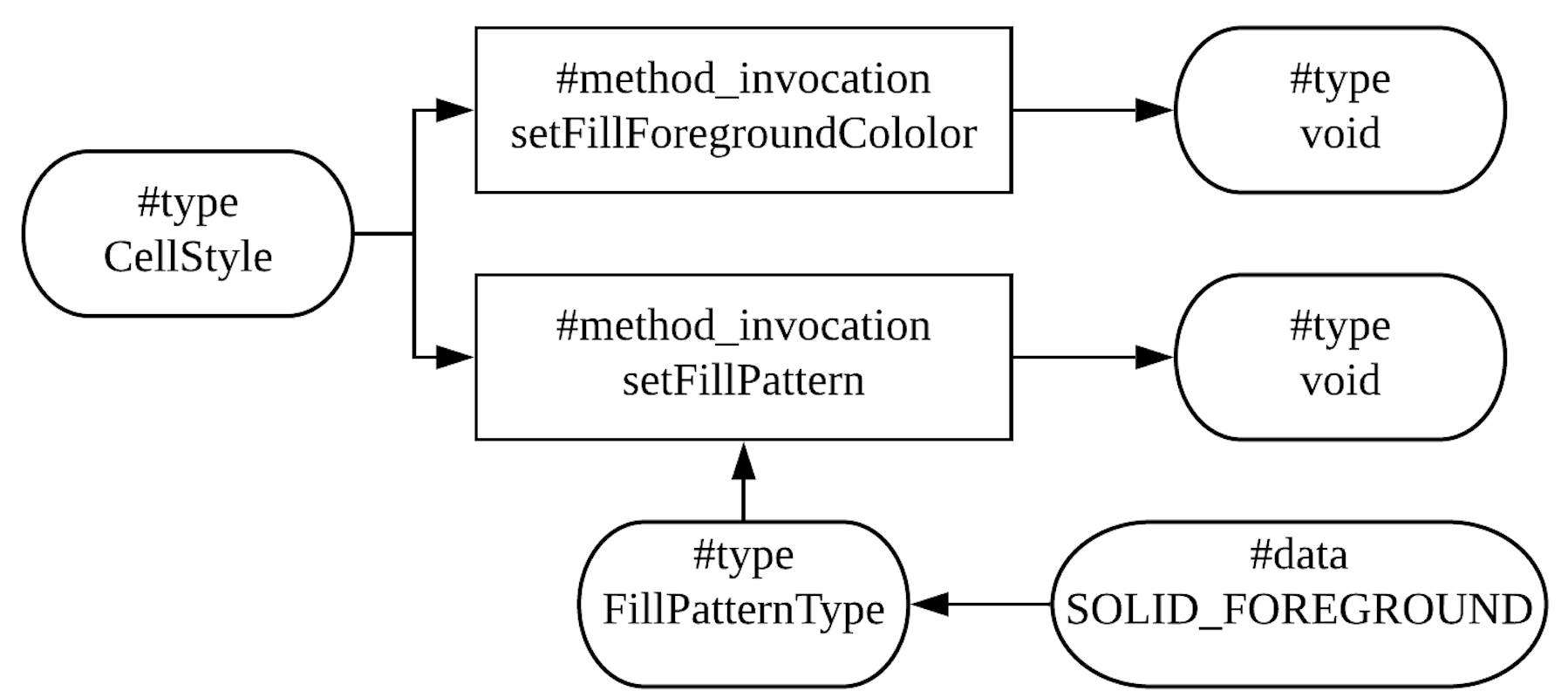}
    \caption{An example data flow graph with type annotations}
    \label{fig-dataflow}
\end{figure}

\begin{figure*}[htb]
    \centering
    \includegraphics[width=0.9\textwidth]{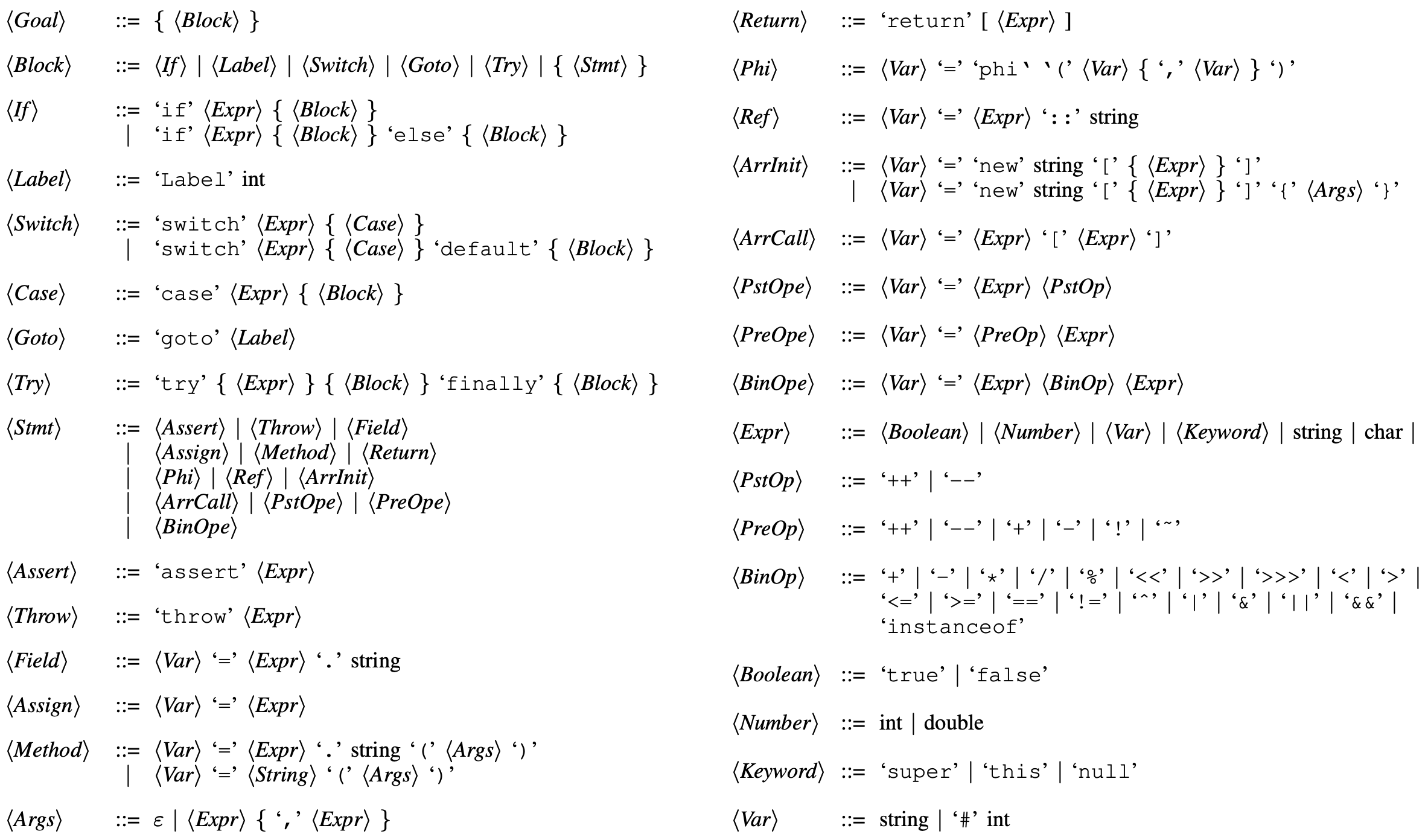}
    \caption{Grammar of intermediate representation for Java code}
    \label{fig-bnf}
\end{figure*}

Source code can be viewed as plain text, however, such simple representation is sensitive to trivial differences (\textit{e.g.}, variable names, indentations) and affects the performance of pattern mining.
Thus, before applying the frequent pattern mining algorithm on the constructed code corpus, we need to abstract the code into a certain data structure.
Common abstractions include the AST (abstract syntax tree) and the method call sequence.
As the natural representation of source code, the AST is sensitive to the coding style differences (\textit{e.g.}, using different keywords \textit{``for''} and \textit{``while''} to implement loops).
Recently, \citep{swim, api-misuse} applied the \textit{structured call sequence} as the code abstraction to mine API-centric code patterns.
The sequence model allows users to define the interested parts of code such as API invocations and guard conditions.
However, sometimes changing the order of certain API calls does not affect the program behavior.
For example, the two snippets in Figure \ref{poi-order} behave in the same way.
In this case, the sequence model is too sensitive to capture the complete pattern.

Compared to the AST and method call sequence, the graph model is more expressive to describe interactions between variables \citep{nguyen2009graph}.
In this paper, we augment the data flow graphs by annotating the data nodes with API types.
Vertices in a data flow graph can be divided into data and operations.
To better fit the library reuse problem, we annotate each data node with the corresponding API type name.
Figure \ref{fig-dataflow} displays the same dataflow graph generated for the two snippets from Figure \ref{poi-order}.
The annotations \textit{``CellStyle''} and \textit{``FillPatternType''} are API types from the library \texttt{apache-poi}.
Also, the different order of method invocations did not affect their abstractions because they share the same data flow.

We follow the common workflow to generate data flow graphs from source code.
First, we generate a self-designed intermediate representation (IR) from Java code.
The IR is independent of the source language and is designed to be conducive for further processing.
Second, we generate control flow graphs from the IR, furthermore, the graphs are refined into the static single assignment (SSA) form.
Third, the control flow graphs are transformed into the data flow graphs.
The last two steps are the implementation of existing algorithms \citep{DBLP:conf/cc/BraunBHLMZ13} and won’t be discussed here.
The rest of this subsection will discuss our self-designed IR, which is shown in Figure \ref{fig-bnf}.

\begin{figure}[htb]
\lstset{language=Java}
\begin{lstlisting}
// for-each style iteration
for (String s: lst) {
    cnt++; foo(cnt, s);
}
// iterator style iteration
Iterator<String> iter = lst.iterator();
while (iter.hasNext()) {
    cnt += 1; foo(cnt, iter.next());
}
\end{lstlisting}
\caption{Two example snippets of different coding styles}
\label{fig-codingstyle}
\end{figure}

There are two reasons to design our own intermediate representation:
First, most existing tools to generate Java IR behave poorly on the incomplete code snippets.
\textit{e.g.}, The famous tool \textsc{Soot}\footnote{https://github.com/Sable/soot} requires all dependencies of the current file to generate the corresponding intermediate code.
While our tool only requires that the input snippets can be taken as a compilation unit, which can be a method without the wrapper class, or even just a block containing several method invocations.
Second, the syntax of Java is complex, there are multiple ways to write code sharing the same behavior.
As Figure \ref{fig-codingstyle} shows, to increments a variable, one can write \emph{cnt++} or \emph{cnt += 1}.
To iterate a list of strings, a for-each loop or an iterator are both correct.
Such details have not been normalized in existing tools, while our IR can eliminate some common coding style differences.
As a result, the two snippets will result in the same representation in our intermediate code.
To be specific, both increment operations are represented by $\langle PstOp\rangle$`++' defined in Figure \ref{fig-bnf}.

After generating data flow graphs for the corpus, we apply gSpan algorithm again to mine frequent subgraphs as code patterns.

\subsection{Skeleton Code}
\begin{figure}[htb]
    \centering
    \includegraphics[width=0.45\textwidth]{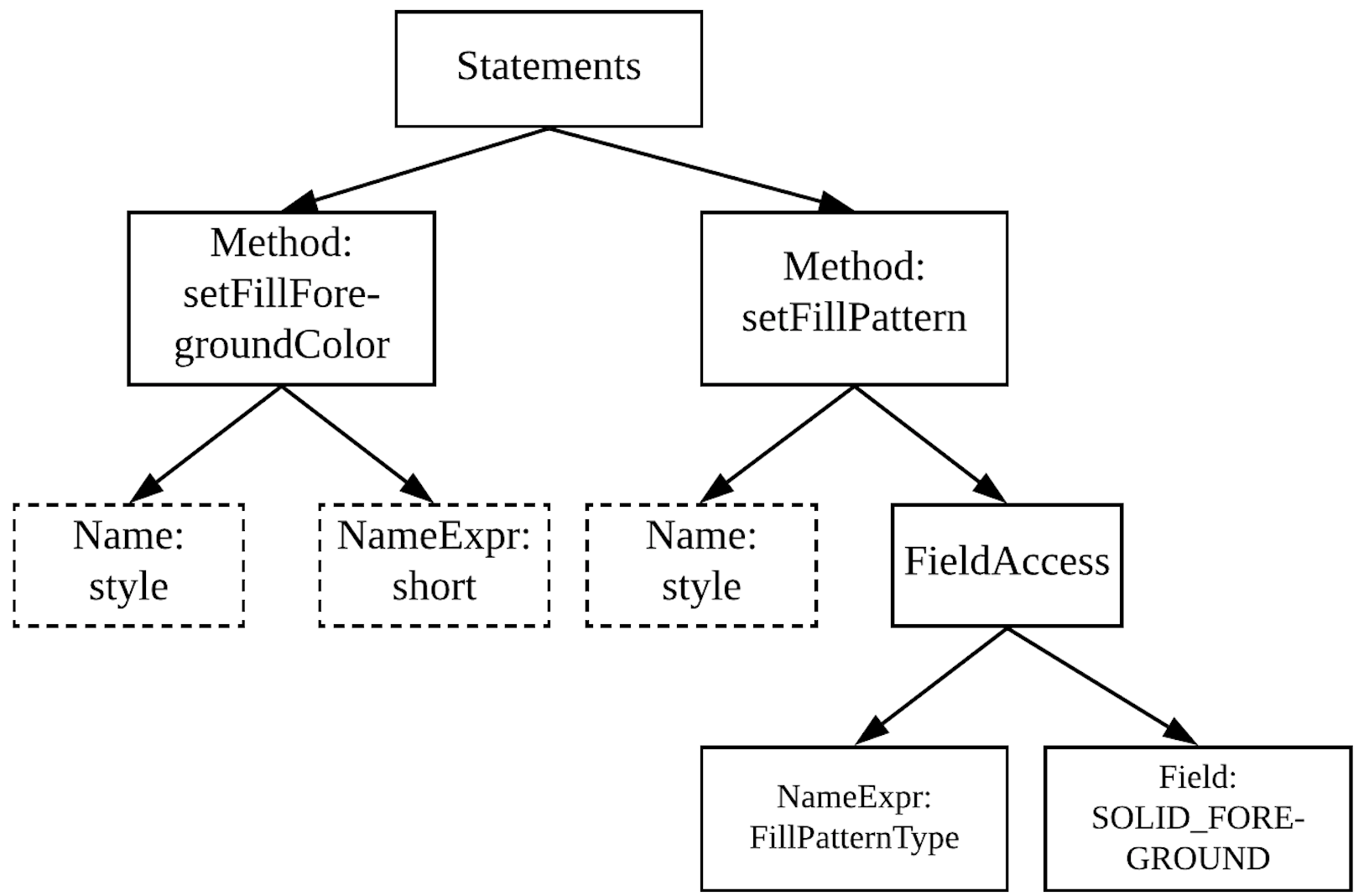}
    \caption{Example skeleton code}
    \label{fig-skeletontree}
\end{figure}

As the last step of pattern mining, we recover the graph-form code patterns into the skeleton code.
\begin{mydef}
    Skeleton code is an incomplete syntax tree, which is obtained by removing trees rooted at $v_1, v_2, ..., v_n$ from a complete syntax tree.
    Each $v_i$ is a node from the complete syntax tree and we name such nodes as holes in the skeleton code.
\end{mydef}
For example, Figure \ref{fig-skeletontree} is the skeleton code recovered from the graph in Figure \ref{fig-dataflow}.
Nodes wrapped in the dotted line are holes in the syntax tree.
Figure \ref{fig-skeleton} shows the text form of the skeleton code.

\begin{figure}[htb]
\lstset{language=Java}
\begin{lstlisting}
<HOLE1>.setFillForegroundColor(<HOLE2>);
FillPatternType fillPatternType1 = 
    FillPatternType.SOLOD_FOREGROUND;
<HOLE1>.setFillPattern(fillPatternType1);
\end{lstlisting}
\caption{Text-form of the example skeleton code}
\label{fig-skeleton}
\end{figure}

During the generation of data flow graphs, we record the corresponding nodes from the syntax tree.
To construct the skeleton code from a data flow graph, we first list all the tree nodes included in the graph.
Then we randomly select a syntax tree of the original source code and search the least common ancestor (LCA) of the nodes in the tree.
After the search, we recover a complete syntax tree containing all the nodes from the graph.
Naturally, the missing parts (\textit{i.e.}, nodes not covered by the graph) in the recovered syntax tree become holes in the skeleton code.
\section{Synthesizer}
\label{synthesizer}
As the last component of our framework, the synthesizer completes the skeleton code into a well-typed code snippet under the current programming context.
We explain the details in this section for reproducibility, but we do not claim the synthesizer as a contribution.

Consider each hole in the skeleton code is annotated with the corresponding type, the synthesis problem can be stated as: 
given a programming context, how to create an expression with the desired type $\tau$.
Here are the three strategies we use:
\begin{itemize}
\item pick a variable of $\tau$ from the current context
\item call the constructor function of $\tau$ 
\item invoke a method chain and the return type of the last method is $\tau$ 
\end{itemize}

\begin{figure}[htb]
    \centering
    \includegraphics[width=0.42\textwidth]{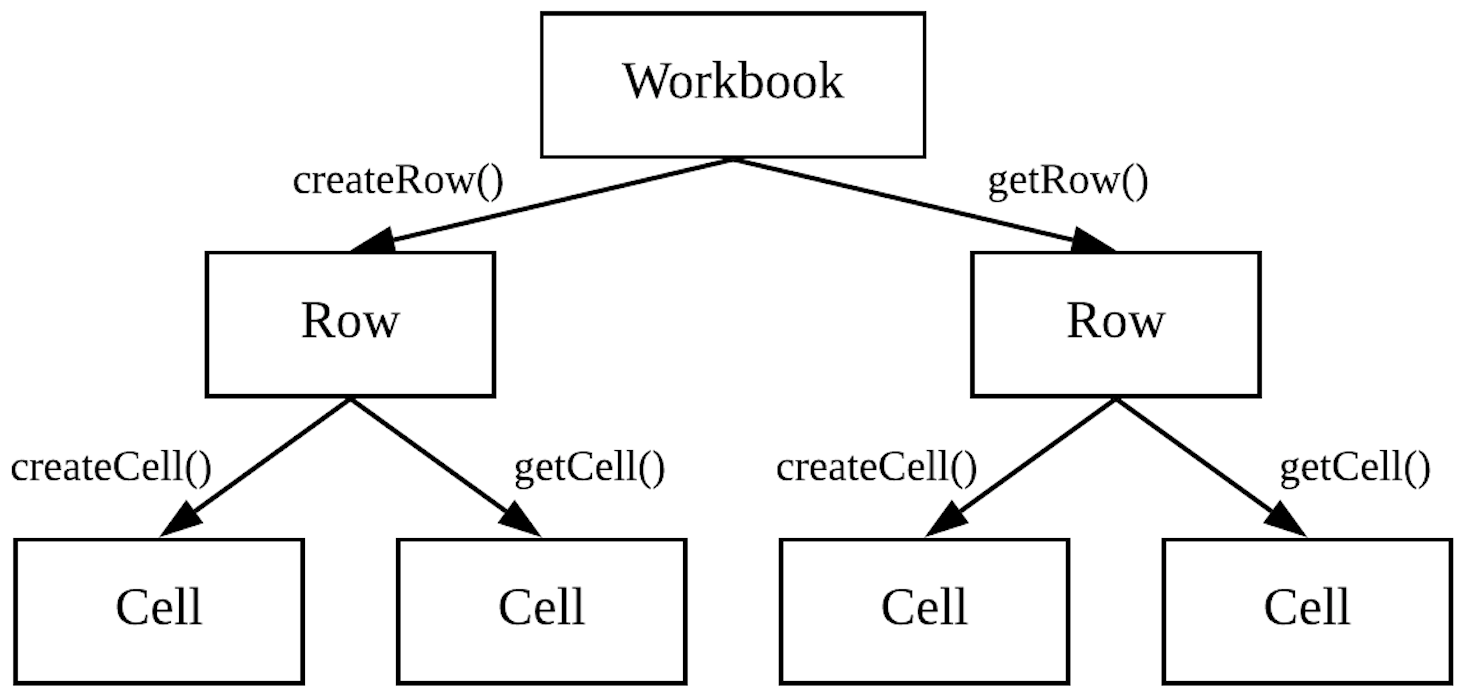}
    \caption{Type-directed search tree}
    \label{search-tree}
\end{figure}

The last strategy is a search process. Figure \ref{search-tree} displays an example of the search tree.
Each node in the tree is an API type from the library and an edge connects two nodes if they are separately the caller and the return type of a method.
The root of the tree is the type of a declared variable and the leaves are the target type.
A path from the root to a leaf represents a method chain which returns the desired type.
As Figure \ref{search-tree} shows, there are four method chains to create a variable with type \textit{``Cell''} from the starting type \textit{``Workbook''}.
During the search, we also considered type casting between types by analyzing the inheritance between APIs.

To guide the search process, we define a cost model as the heuristic rule.
The model evaluates the goodness of different ways for variable synthesis by mapping them to integers.
Using existing variables in context is encouraged, with zero cost.
If there are multiple variables with the same type, we choose the one created most recently due to software localness.
If a variable is the return value of a certain method, it costs 2 when the method is a constructor and 1 for else.
The process for variable synthesis could be recursive, which means in the process of synthesizing the current variable, the invocations require parameters that are not in the context.
Our cost model adds the costs for synthesizing these parameters to the total cost.

\begin{equation}
\begin{aligned}
cost(t) = 0,\quad t\ in\ context\ or\ t\ is\ constant
\end{aligned}
\end{equation}
\begin{equation}
\begin{aligned}
cost(f(t_1, t_2, ..., t_k)) =
price(f) + \sum_{i=1}^{k}{cost(t_i)}
\end{aligned}
\end{equation}
\begin{equation}
\begin{aligned}
price(f) =
\begin{cases}
2& \text{f is constructor}\\
1& \text{else}
\end{cases}
\end{aligned}
\end{equation}

If two expressions get the same score under the cost model, we break the tie by referring to the code corpus. 
Recall that each skeleton code comes from a code corpus, we first select instances of the skeleton code from the corpus.
For each instance, we extract the variable to fill the hole and locate the definition of the variable by analyzing the \textit{``def-use''} relationship.
The process of extraction can be recursive because the definition of a variable may use other undeclared variables.
The recursion terminates when we find all definitions of the variables or we meet a variable defined outside the current method body (\textit{i.e.}, parameters of the method, global variables).
As a result, from each instance in the corpus, we extract an expression (\textit{i.e.}, a method chain) to fill the hole. 
For the two synthesized expressions with the same score, we calculate their frequency in the expressions extracted from the corpus and recommend them in the order of decreasing frequency.
\section{Evaluation}
\label{evaluation}

In this section, we evaluate \textsc{NLI4J} from three perspectives, corresponding to the three components of the framework.
First, to evaluate the accuracy and completeness of the functional features, we compare the extracted functional features with the library functionalities provided by the official tutorials.
Second, to evaluate the quality of code patterns, we use the code examples from the official tutorials as benchmarks and compare our mining algorithm with two existing pattern mining tools \citep{OOPSLA15, api-misuse}.
Third, we evaluate the synthesizer with a controlled experiment on \texttt{apache-poi}.
In the study, we implement an IDE plugin by putting all the three components together and investigate whether the plugin could save programmers' time to solve real-world tasks.

Our research questions are as follows:
\begin{itemize}
    \item \textit{$RQ_1:$ How well does our filtering pipeline perform on selecting functional features from user discussions?}
    This question aims at accessing whether our filtering pipeline is effective to filter unrelated verb phrases.
    Furthermore, we investigate the importance of each filter in the process.
    \item \textit{$RQ_2:$ To what extent is \textsc{NLI4j} able to provide accurate and complete functional features?}
    This question evaluates the accuracy and completeness of the normalized functional features.
    Here, accuracy refers to each functional feature should clearly correspond to a functionality.
    The completeness refers to the capability of the generated functional features to cover frequently-used library functionalities.
    \item \textit{$RQ_3:$ How does our code pattern mining algorithm perform compared to existing mining tools?}
    This research question is related to the quality of the mined code patterns.
    Given the same codebase, we compare our mined code patterns with two existing mining algorithms, which separately abstract source code into syntax trees and sequences.
    \item \textit{$RQ_4:$ To what extent is \textsc{NLI4j} able to promote the efficiency to solve real-world programming tasks?}
    Finally, this research question directly investigates the usefulness of \textsc{NLI4j} in real-world development.

\end{itemize}

In the following, we first introduce our datasets and benchmarks.
Then, for each research question, we detail our evaluation methodology and results in an individual subsection.

\subsection{Datasets and Benchmarks}
To answer the research questions, we collect data for five Java libraries:
an html extraction library (\texttt{jsoup}), a source code parser (\texttt{eclipse-jdt}), 
a library manipulating Microsoft documents (\texttt{apache-poi}), a deep learning toolkit (\texttt{deeplearning4j}) and a graph database platform (\texttt{neo4j}).
In addition to being widely used, these five libraries cover different domains of programming, from the front-end html parsing to the back-end database manipulation.

To construct NLI for a given library, our tool requires 1). related threads from Stack Overflow, and 2). client code reusing the library APIs.

Stack Overflow provides a tag for each of the five libraries (\textit{e.g.} tag \textit{``jsoup''} for the \texttt{jsoup} library).
For each library, we crawl all the threads containing tag \textit{``java''} and the library-specific tag.
Since our functional feature extractor processes a single sentence at a time, we extract textual contents of the threads and split the text into sentences using the Stanford NLP toolkit.
The sentences form our first dataset \textit{$SO_{large}$}.
Table \ref{so-dataset} lists the number of the threads and the split sentences in \textit{$SO_{large}$}.
Furthermore, we extract a smaller dataset \textit{$SO_{small}$} by randomly sampling 100 sentences for each of the five libraries.
During the sampling, we remove sentences that are shorter than 15 characters, since such sentences are usually mistakenly split and seldom contain functional features.
As a result, the dataset \textit{$SO_{small}$} contains 500 sentences.
Based on our theoretical definition of the extraction process, the first author manually labels the functional features for each sentence in \textit{$SO_{small}$}.

For client code, we build the dataset by downloading all the client repositories using the Github APIs \footnote{https://api.github.com/search/repositories}.
Given a library, the query we used is restricted as follows: the body is the name of the library (\textit{e.g.}, jsoup), the programming language is specified as Java, and each repository should have at least five stars.
Table \ref{client-dataset} lists the number of the client repositories we download and the number of the source files from the repositories.

To evaluate the generated NLI, a list of library functionalities and their implementations are required as benchmarks.
We turn to the official tutorial for each of the five libraries.
The names of the tutorials vary between libraries, (\textit{e.g.} cookbook, developers' guide), and we organize each tutorial as a list of functionalities.
Each functionality is a pair consisting of a concise description and a code example.
We filter the functionalities with too long code examples (\textit{i.e.} more than 20 lines of code after removing the comments) because instead of discussing a specific feature, such long examples are more likely to describe a topic or a complete procedure to reuse the library.
After the filtering, we treat all the left official functionalities as benchmarks in our evaluation.
For each library, Table \ref{benchmarks} lists the number of the functionalities and the average lines of a code example (LoC) in the benchmarks.

\begin{table}
    \centering \caption{Overview of the Stack Overflow dataset}
    \vspace{0.2cm}
    \label{so-dataset}
    \begin{tabular}{l r r}
        \hline
        \textbf{\footnotesize Library} & \textbf{\footnotesize \# Threads} & \textbf{\footnotesize \# Sentences}\\
        \hline
        jsoup & 649 & 2,780\\
        apache-poi & 2,496 & 8,046\\
        neo4j & 1,600 & 8,144\\
        deeplearning4j & 290 & 1,310\\
        eclipse-jdt & 805 & 3,461\\
        \hline
        \bfseries all & \bfseries 5,840 & \bfseries 23,741\\
        \hline
    \end{tabular}
\end{table}

\begin{table}
    \centering
    \caption{Overview of the client code dataset}
    \vspace{0.2cm}
    \label{client-dataset}
    \begin{tabular}{l r r}
        \hline
        \textbf{\footnotesize Library} & \textbf{\footnotesize \# Repositories} & \textbf{\footnotesize \# Source Files}\\
        \hline
        jsoup & 119 & 5,077\\
        apache-poi & 239 & 21,601\\
        neo4j & 291 & 37,428\\
        deeplearning4j & 48 & 7,470\\
        eclipse-jdt & 26 & 34,254\\
        \hline
        \bfseries all & \bfseries 723& \bfseries 105,830\\
        \hline
    \end{tabular}
\end{table}

\begin{table}[!htb]
    \centering
    \caption{Benchmarks from the offcial tutorials}
    \vspace{0.2cm}
    \label{benchmarks}
    \begin{tabular}{l r r}
        \hline
        \textbf{\footnotesize Library} & \textbf{\footnotesize \# Functionalities} & \textbf{\footnotesize Average LoC}\\
        \hline
        jsoup & 13 & 4.1\\
        apache-poi & 46 & 12.3\\
        neo4j & 9 & 2.9 \\
        deeplearning4j & 21 & 10.9\\
        eclipse-jdt & 6 & 18.0\\
        \hline
        \bfseries all & \bfseries 95 & \bfseries 10.3\\
        \hline
    \end{tabular}
\end{table}
\subsection{RQ1: Filtering Pipeline}
As the first step of our algorithm, \textsc{NLI4j} extracts all the verb phrases from user discussions and then select functional features by filtering unrelated phrases.
This subsection discusses the output of our filtering pipeline on the labeled dataset \textit{$SO_{small}$}.

\subsubsection{Methodology}
We first combine all the three filters (\textit{i.e.} stop word filter, context filter, and structure filter) to filter verb phrases.
The results are compared to the manually labeled results on the dataset \textit{$SO_{small}$}.
We automatically compare the results with a script that simply matches the textual contents.
To avoid the mistakes caused by trivial details in natural language (\textit{e.g.} tenses of verbs, the plural form), instead of asking the annotator to label the benchmarks from scratch, 
we provide the extracted verb phrases and let the annotator select the functional features from the phrases.
If the provided phrases already missed certain functional features, the verdict of this sentence will be a failure even before comparison. 

Furthermore, to evaluate the importance of each filter, we created three new filtering pipelines by removing one filter at a time.
Then, we evaluated the three modified pipelines using the same script.

\subsubsection{Results}
From the 500 sentences, our extractor extracted 1,360 verb phrases using the Stanford NLP toolkit.
We fed the phrases to our filter pipeline and got 315 functional features.
For 93\% (465 out of 500) sentences, the automatically extracted functional features matched the labeled ones in the benchmark.
Our tool missed 41 functional features and gave 12 wrong features in the remaining 35 sentences.
Table \ref{label-result} summarizes details of the results for each library.
For each library, we list 1). the number of verb phrases mined from the sentences, 2). the number of functional features after filtering, and 3). the number of sentences which the filtered results match the benchmarks.
From the result, we did not notice significant variations between different libraries.
However, it is possible that our fixed stop words can cause some false negatives when a new library is specified, since a stop word could be a domain-specific concept or action for the new library.
In that case, a customized stop word list is recommended.

\begin{table}
    \centering
    \caption{Filtering results for the five libraries}
    \vspace{0.2cm}
    \label{label-result}
    \begin{tabular}{l  r  r  r}
        \hline
        \multirow{2}*{\textbf{\footnotesize Library}} & \multirow{2}*{\textbf{\footnotesize \# Phrases}} & \multirow{2}*{\textbf{\footnotesize \# Features}} & \textbf{\footnotesize \# Correct} \\ 
        & & & \textbf{\footnotesize Sentences} \\
        \hline
        jsoup & 245 & 55 & 96 \\
        apache-poi & 277 & 81 & 92 \\
        neo4j & 290 & 75 & 87 \\
        deeplearning4j & 284 & 68 & 96 \\
        eclipse-jdt & 264 & 36 & 94 \\
        \hline
        \bfseries all & \bfseries 1,360 & \bfseries 315 & \bfseries 465 \\
        \hline
    \end{tabular}
\end{table}

We checked each of the 35 failed sentences and summarized two main reasons for the mistakes, which resulted in both missing and wrong functional features:
\begin{itemize}
    \item \emph{Preprocessing of natural language.}
    We found in more than half of the failed sentences, the NLP toolkit did not produce the correct verb phrase list as expected.
    \item \emph{Tangled votes.}
    Some phrases were upvoted and downvoted at the same time, and our current weights for the filters lead to a wrong decision for these phrases.
\end{itemize}

The first reason is an external factor to our tool.
We found the Stanford NLP toolkit sometimes failed to split the sentences correctly when the punctuation characters are not correctly used.
Also, a common case for failed POS tagging is when verbs appear at the beginning of sentences.
For the second reason, as there were both upvotes and downvotes in our filter pipeline, sometimes they are tangled and bring mistakes in the functional feature recognition.
For example, the phrase \emph{``return the node of the highest score''} was missing from the sentence \emph{``With Cypher, I'm trying to return the node of the highest score."} because it was downvoted for using a stop word \emph{return} and upvoted for the context (with a preceding Q\&A expression \emph{I'd like to}).
Machine learning approaches could help in such a scenario by assigning proper weights for the three filters (all one point in our current implementation).
However, consider the small size of the annotated sentences and the fact that the current algorithm is accurate for most sentences, we did not apply machine learning approaches at present.

To evaluate the importance of each filter, we create three new filtering pipelines by removing one filter at a time.
The results are displayed in the first three rows of Table \ref{tab-modified} and the last row combines all the three filters.
When the stop word filter is removed, the number of wrong features rapidly raises to 478 as the first row shows.
The context filter upvotes the verb phrases following Q\&A expressions, as a result, the filter pipeline tends to give a lower score for each phrase after removing this filter.
As the second row shows, the number of missing features without the context filter is the largest.
The third row depicts the result for removing the structure filter, which also brings more incorrect features.

\begin{table}
    \centering
    \caption{Results for different combinations of the three filters}
    \label{tab-modified}
    \vspace{0.2cm}
    \begin{tabular}{l  r  r  r}
        \hline
        \textbf{\footnotesize Filter} & \textbf{\footnotesize \#Correct} & \textbf{\footnotesize \#Wrong} & \textbf{\footnotesize \#Missing}\\
        \textbf{\footnotesize Combinition} & \textbf{\footnotesize Sentences} & \textbf{\footnotesize Features} & \textbf{\footnotesize Features}\\
        \hline
        context+structure & 120 & 478 & 21 \\
        word+structure & 414 & 10 & 103 \\
        word+context & 389 & 97 & 36 \\
        \hline
        all filters & 465 & 12 & 41\\
        \hline
    \end{tabular}
\end{table}

\mybox{\emph{Answer for RQ1:} 
On a labeled dataset containing five hundred sentences, our filtering pipeline correctly filters
 unrelated verb phrases for 465 (93\%) sentences. All the three filters contribute to the performance
 and the stop word filter is proved to be the vital factor. }

\subsection{RQ2: Functional Features}
In this subsection, we generate functional features for each library from the dataset \textit{$SO_{large}$}.
Our benchmark is the lists of functionality descriptions from the official tutorials.
The evaluation checks both 1). whether each functional feature is accurate, and 2). whether each functionality in the tutorial is covered.

\subsubsection{Methodology}
Given the Stack Overflow corpus of a library, the output of our functional feature extractor is a list of functional features in verb phrase form.

To evaluate the accuracy, we ask two annotators to rate the extracted functional features.
They are requested to give a score for each feature: two points for an actual library functionality, one point for a likely functionality that requires further information to make it clear, zero point for a meaningless phrase.
Two annotators mark the functional features separately and afterwards discuss to reach an agreement.
We count all the ratings by annotators and calculate the average score for all the functional features.

To evaluate the completeness, we ask the annotators to review the functionalities in the benchmark one by one and judge whether the functionality is included in our generated functional features.
Again, we ask the annotators to give a score for each functionality.
If a functionality is included in our generated features, our result gets two points.
If our output includes a similar functional feature but not precise, our result gets one point.
Otherwise, our result gets zero point for the functionality.

\subsubsection{Results}

Table \ref{tab-task1} displays the results for the accuracy of the functional features.
The first column is the name of the library and the second column is the number of the normalized functional features extracted from the \textit{$SO_{large}$} dataset.
The third column lists the number of functional features marked with three different scores and the last column is the average score for all the functional features in this library.
As the last row shows, for a total of 531 functional features, 282 (53.1\%) of them are annotated with two points, 176 (33.1\%) are annotated with 1 point and the remaining 73 (13.7\%) are irrelevant to library functionalities.
The average score shows that our functional features get approximately 1.39 points out of two.

Table \ref{tab-task2} displays the results for the completeness of the functional features.
Instead of rating a functional feature, we rate each functionality from the official tutorial in Table \ref{tab-task2}.
The result shows that our generated functional features can cover 86.3\% (82 out of 95, 66.3\% with two points, 20.0\% with one point) of the functionalities in the benchmark.
For the functions which get one point, our annotators reported that the majority of them are caused by the fact that the tutorial summarizes several tasks into one functionality.
For example, the last function in the tutorial of \texttt{apache-poi} is \textit{``cells with multiple styles''}, which mentioned three tasks (setting color, font and cell style) at the same time.
There is little chance that a user will discuss the three functionalities together in a verb phrase.
We carefully analyzed all the 13 missing functionalities (rated as zero point) by manually searching them on Stack Overflow.
As a result, seven of them are never mentioned on Stack Overflow, the rest six functionalities are discussed fewer than three times in the whole corpus.
Since we only keep the frequent normalized syntax trees during normalizing functional features, phrases with low frequency will not be included in our final result.

The results show a little fluctuation among different libraries, especially for the completeness.
In Table \ref{tab-task2}, the \texttt{eclipse-jdt} library gets the highest score of 2.00, while the \texttt{deeplearning4j} library is rated the lowest (1.09 points).
The fluctuation comes from the different sizes of related threads on Stack Overflow.
In fact, as Table \ref{so-dataset}, the number of threads under the tag \texttt{deeplearning4j} is the smallest in our dataset.
The small size of discussions obviously affects the completeness of functional features, which is an external threat to our algorithm.

\begin{table}[t]  
    \centering
    \caption{Accuracy of the generated functional features}  
    \vspace{0.2cm}
    \label{tab-task1}
    \begin{tabular}{l c p{0.5cm}<{\centering} p{0.5cm}<{\centering} p{0.5cm}<{\centering} c c}
    \hline
    \multirow{2}*{\textbf{\footnotesize Library}} & \multirow{2}*{\textbf{\footnotesize \# Features}} & \multicolumn{3}{c}{\textbf{\footnotesize Score}} & \multirow{2}*{\textbf{\footnotesize Average}} \\ 
    \cline{3-5}
    & & 2 & 1 & 0 & \textbf{\footnotesize Score} \\
    \hline 
    jsoup & 86 & 41 & 27 & 18 & 1.26\\
    apache-poi & 190 & 116 & 48 & 26 & 1.47\\
    neo4j & 119 & 61 & 40 & 18 & 1.36\\
    dl4j & 33 & 16 & 14 & 3 & 1.39\\
    eclipse-jdt & 103 & 48 & 47 & 8 & 1.39\\
    \hline
    {\bfseries all} & {\bfseries 531} & {\bfseries 282} & {\bfseries 176} & {\bfseries 73} & {\bfseries 1.39} \\
    \hline
    \end{tabular}
\end{table}

\begin{table}[t]  
\centering
\caption{Completeness of the generated functional features}
\vspace{0.2cm}
\label{tab-task2}
\begin{tabular}{l c p{0.5cm}<{\centering} p{0.5cm}<{\centering} p{0.5cm}<{\centering} c}
    \hline
    \multirow{2}*{\textbf{\footnotesize Library}} & \multirow{2}*{\textbf{\footnotesize \# Functions}} & \multicolumn{3}{c}{\textbf{\footnotesize Score}} & \multirow{2}*{\textbf{\footnotesize Average}} \\ 
    \cline{3-5}
    & & 2 & 1 & 0 & \textbf{\footnotesize Score} \\
    \hline
    jsoup & 13 & 10 & 3 & 0 & 1.77 \\
    apache-poi & 46 & 30 & 12 & 4 & 1.56\\
    neo4j & 9 & 7 & 1 & 1 & 1.67 \\
    dl4j & 21 & 10 & 3 & 8 & 1.09 \\
    eclipse-jdt & 6 & 6 & 0 & 0 & 2.00\\
    \hline
    {\bfseries all} & {\bfseries 95} & {\bfseries 63} & {\bfseries 19} & {\bfseries 13} & {\bfseries 1.52} \\
    \hline
\end{tabular}
\end{table}

\mybox{\emph{Answer for RQ2:} 
By comparing the functional features with the official tutorials,
 we found that 86.2\% (458 out of 531) functional features are accurate.
 Furthermore, the features can cover 86.3\%(82 out of 95) of the functionalities listed in the official tutorials.
}

\subsection{RQ3: Code Pattern}
In this section, we focus on evaluating the performance of our code pattern mining algorithm.
We start the evaluation from some examples and then compare our miner with two existing pattern mining tools.

\subsubsection{Examples}

Figure \ref{fig-patternexp} shows five code patterns that \textsc{NLI4j} mines.
A symbol with $\$$ denotes that there is a missing part in the code pattern.
To be more specific, a $<\$HOLE>$ represents a missing variable and a $<\$BODY>$ represents a missing code block.
The reader will observe the immediate usefulness of the code patterns for learning API usage.

\begin{figure}[t]
\begin{lstlisting}[language=Java]
// parse text from html
Document document_1 = Jsoup.parse(<$HOLE1>);
document_1.select(<$HOLE2>).first().text();
\end{lstlisting}

\begin{lstlisting}[language=Java]
// create an embedded database
GraphDatabaseFactory factory_1 = 
    new GraphDatabaseFactory();
GraphDatabaseService service_1 = 
    factory_1.newEmbeddedDatabase(<$HOLE1>);
\end{lstlisting}

\begin{lstlisting}[language=Java]
// configure a network
MultiLayerConfiguration configuration_1 =
    new NeuralNetConfiguration.Builder()
    .seed(<$HOLE1>).iterations(<$HOLE2>)
    .list().layer(<$HOLE3>).build();
\end{lstlisting}

\begin{lstlisting}[language=Java]
// merge cells
CellRangeAddress address_1 = new
    CellRangeAddress(
    <$HOLE1>, <$HOLE2>,
    <$HOLE3>, <$HOLE4>
);
<$HOLE5>.addMergedRegion(address_1);
\end{lstlisting}

\begin{lstlisting}[language=Java]
// save workbook
Workbook wb_1 = new HSSFWorkbook();
try {
    wb.write(<$HOLE1>);
} catch (IOException e) { <$BODY> }
\end{lstlisting}

\caption{Example code patterns for functional features\label{fig-patternexp}}
\end{figure}

As Figure \ref{fig-patternexp} shows, a code pattern usually describes the frequent combination of API elements.
The fourth code pattern (\textit{i.e., ``merge cells''}) denotes that a cell region is managed by the class $CellRangeAddress$, which is usually invoked with another method $addMergedRegion$.
To instantiate a $CellRangeAddress$ object, four parameters are required to specify the left top and the right bottom corners of the region.
Some patterns contain control flow statements besides API invocations, such as the last example \textit{``save workbook''} in Figure \ref{fig-patternexp}.
The code pattern not only summarizes the correct APIs to invoke, but also hints that the method $write$ needs to handle an exception.
The specific way to handle the exception is left to the user in a $<\$BODY>$ block.

\subsubsection{Methodology}

To evaluate the performance of our pattern mining algorithm, we use code examples from the official tutorials as the benchmark.
From the total 95 functionalities in the benchmarks, we first remove the 13 functionalities that are not covered by our functional features (\textit{i.e.} the functionalities with zero point in Table \ref{tab-task2}).
Then we removed another 13 functionalities because our algorithm failed to match a correct API from the corresponding functional feature.
Table \ref{pattern-dataset} shows the 69 left functionalities.

\begin{table}[htb]
    \centering
    \caption{The number of functionalities to mine code patterns}
    \vspace{0.2cm}
    \label{pattern-dataset}
    \begin{tabular}{l r r}
        \hline
        \multirow{2}*{\textbf{\footnotesize Library}} & \textbf{\footnotesize \# Covered} & \textbf{\footnotesize \# Functions to}\\ 
        & \textbf{\footnotesize Functions} & \textbf{\footnotesize Mine Patterns} \\
        \hline
        jsoup & 13 & 12\\
        apache-poi & 42 & 36\\
        neo4j & 8 & 7\\
        deeplearning4j & 13 & 10\\
        eclipse-jdt & 6 & 4\\
        \hline
        \bfseries all & \bfseries 82 & \bfseries 69\\
        \hline
    \end{tabular}
\end{table}

As there is no universal metric to measure the quality of code patterns, we approximated the quality by calculating the \emph{Jaccard distance}.
To be more specific, we built a set of the invoked APIs in the code pattern and another set of the invoked APIs in the official example.
Jaccard distance is the metric to calculate the differences between two sets as follows:
\begin{equation}
dis(X,Y)=1-\frac{|X\cap Y|}{|X\cup Y|}
\end{equation}

We compare our pattern mining algorithm with two existing tools, \textit{i.e.}, \textsc{anyCode} and \textsc{ExampleCheck}.
\textsc{anyCode} expands an API element into a Java expression with a pre-trained PCFG (Probabilistic Context Free Grammar) model.
The second tool \textsc{ExampleCheck} is designed to check API misuse from the Stack Overflow.
The rationale behind is to compare API examples from Stack Overflow with the mined API usage patterns from Github.
We choose the two tools because their abstractions for source code are representative. 
\textsc{anyCode} abstracts code into tree-based structure PCFG, and \textsc{ExampleCheck} abstracts code into the sequence structure SCS (\textit{i.e.}, Structured Call Sequence).
To make the comparison meaningful, we configure the settings for all three tools as follows:
\begin{itemize}
    \item \textit{The same codebase.}
    All three tools are provided with the same codebase, which is all the usage examples for a given API. On average, the codebase for each API contains 217 source code files.
    \item \textit{The same threshold.}
    We set the threshold (5\%) for the minimum frequency of a pattern to be mined from the codebase.
\end{itemize}

\subsubsection{Results}
Table \ref{pattern-result} shows the results of the experiments.
Given a tool and a library, we list the average Jaccard distance between the mined code patterns and the code examples from the benchmarks.
Overall, \textsc{NLI4j} achieves the minimum average Jaccard distance (0.29), which proves the code patterns mined by our tool are more similar to the official code examples.
In the experiment, we found that \textsc{anyCode} can only synthesize quite short patterns.
Some code examples in our benchmark contain more than ten API invocations, as a result, the performance of \textsc{anyCode} on these cases are not as good as the other two tools.
\textsc{ExampleCheck} and \textsc{NLI4j} can generate more complete and complex code patterns.
However, as we explained before, in such cases, the sequence structure \textsc{ExampleCheck} used is too strict for pattern mining.
For example, for the color setting task in Apache POI, we found that two APIs (\textit{i.e.}, \textit{``setFillForegroundColor''} and \textit{``setFillPattern''}) could be swapped.
However, swapping two APIs will result in two different subsequences for \textsc{ExampleCheck} and it failed to produce the complete pattern.

Besides, we observed the fluctuation among different libraries.
For \texttt{deeplearning4j} and \texttt{eclipse-jdt}, we found the Jaccard distances of all the three tools are significantly larger than the rest three libraries.
In fact, we found the style of API usage varies among different libraries.
For example, \texttt{deeplearning4j} often requires a long method chain to configure the network from all aspects.
However, users of \texttt{deeplearning4j} may skip some aspects in their client code, as a result, the mined patterns are visibly shorter than the official code examples.
For \texttt{eclipse-jdt}, many functionalities of the library apply the visitor pattern (a design pattern). 
All the code abstractions of the three tools are designed to analyze code snippets inside a method, which could not represent the visitor pattern well.

\begin{table}
    \centering
    \caption{Comparison of three pattern mining tools}
    \vspace{0.2cm}
    \label{pattern-result}
    \begin{tabular}{l  c  c  c}
        \hline
        \textbf{\footnotesize Library} & \textbf{\footnotesize anyCode} & \textbf{\footnotesize ExampleCheck} & \textbf{\footnotesize NLI4j} \\
        \hline
        jsoup & 0.33 & 0.23 & 0.15 \\
        apache-poi & 0.42 & 0.27 & 0.21 \\
        neo4j & 0.35 & 0.29 & 0.29 \\
        deeplearning4j & 0.79 & 0.68 & 0.56 \\
        eclipse-jdt & 0.85 & 0.81 & 0.81 \\
        \hline
        \bfseries Average & \bfseries 0.48 & \bfseries 0.36 & \bfseries 0.29 \\
        \hline
    \end{tabular}
\end{table}

\mybox{\emph{Answer for RQ3:} 
    Given the same codebase, our generated code patterns are more complete and accurate than two existing pattern mining tools.
}

\subsection{Controlled Experiment}
We conducted a controlled experiment on the library \texttt{apache-poi} to see whether our tool can improve the efficiency of reusing libraries in real-world programming.
We also recorded the ranking of the expressions accepted by users to evaluate the performance of our synthesizer.
\subsubsection{Methodology}
\begin{figure*}
\centering
\includegraphics[width=0.95\textwidth]{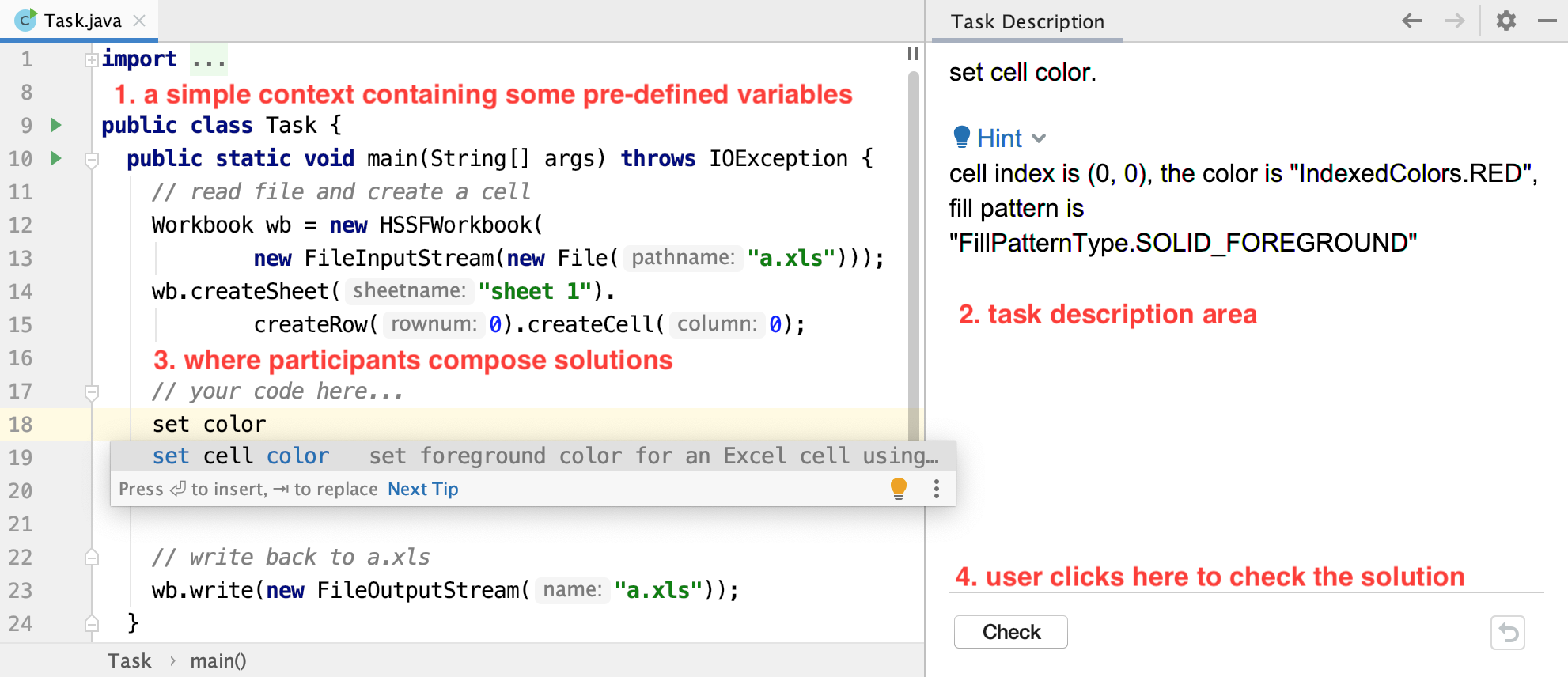}
\caption{UI for our controlled experiment. The user is invoking a functional feature.}
\label{fig-ui}
\end{figure*}
To evaluate the utility and effectiveness of programming with \textsc{NLI4j}, we invited 8 participants to solve real-world programming tasks using the tool.
All the participants were familiar with the Java programming language and were divided into two groups.
The newcomer group consisted of five participants new to the library \texttt{apache-poi}.
The rest three participants once built client projects with \texttt{apache-poi} and they formed the expert group.
The logic and usage of \textsc{NLI4j} were briefly introduced to all the participants in advance.

We prepared five specific programming tasks for the participants to solve.
Prototypes of the tasks were randomly picked from an online tutorial website\footnote{https://www.tutorialspoint.com/apache\_poi}.
We concretized the tasks for two reasons.
First, some tasks require specific configurations to be automatically validated. 
For example, we concretized the task \emph{``create blank workbook''} by specifying the file name and the path to save it.
Second, some tasks in the tutorial are supposed to teach users how to manipulate a class for multiple subtasks. Separating the subtasks makes it more executable for composing solutions and validators.
Table \ref{tab-tasks} lists the tasks with a brief description and the number of API elements invoked in the code example from the tutorial.
On average, one task in our experiment invokes 5.4 API elements.
Figure \ref{fig-ui} shows the user interface of our controlled experiment.
Each task has three components: a task description with a detailed hint, a solution file with some pre-defined variables, and a validator program.
All tasks consider the fill-in-the-blanks approach, which meant the participants needed to fill the solution file by implementing the missing functions.
A task is considered to be accomplished if the validator returns the accepted page.
The tool for our controlled experiment is available in the published online artifacts.

\begin{table}
    \caption{Five tasks for participants to solve with \textsc{Apache-POI}}
    \vspace{0.2cm}
    \begin{tabular}{c|l|c}
    \hline
    \bfseries Id & \bfseries Task description & \bfseries \#invoked APIs\\
    \hline
    1 & create blank workbook & 4\\
    2 & write into a spreadsheet & 3\\
    3 & set cell color & 6\\
    4 & set italic font and font color & 7\\
    5 & create hyperlink to URL & 7\\
    \hline
    \end{tabular}
    \label{tab-tasks}
\end{table}

We allowed all participants to visit online resources such as Q\&A forums and search engines when solving tasks, but we recorded the number of pages they opened in the process.
Two settings were configured for the coding environment, one equipped with the \textsc{NLI4j} plugin and the other without it.
Participants were assigned at random to each programming task and each coding environment, and thus there was no proper balance.
No participants were assigned to the same task with different coding environments.
For the participants who used \textsc{NLI4j}, their interaction with the plugin was recorded.
Recall that our synthesizer would recommend synthesized expressions to users, we recorded whether the user accepted the recommendation and the ranking of the expression that they used.
Finally, the overall task duration and the number of websites viewed was recorded to facilitate data analysis.

\subsubsection{Results}

Table \ref{tab-user} shows the results of the controlled experiment.
Columns (1) and (2) display the participant's index and the reported programming expertise (N stands for the newcomer, and E for the expert).
Columns (3) and (4) display the task and the code environment (STD stands for the standard IDE and NLI stands for the IDE equipped with \textsc{NLI4j} plugin).
Finally, Column (5) refers to the overall duration of the task, and Column (6) displays the number of web pages the participant opened for the task.

\begin{table}[htb]
    \centering
    \scriptsize
    \caption{Summary of experiment results}
    \label{results}
    \centering
    \begin{tabular}{c|c|c|l|c|c}
    \hline
    \bfseries (1) & \bfseries (2) & \bfseries (3) & \bfseries (4) & \bfseries (5) & \bfseries (6)\\
    \bfseries Id & \bfseries Expertise & \bfseries Task & \bfseries Environment & \bfseries Time(s) & \bfseries \#Pages\\
    \hline
    \multirow{5}{*}{1} & \multirow{5}{*}{N} & 1 & NLI & 144 & 0 \\ \cline{3-6} 
    &  & 2 & STD & 377 & 4 \\ \cline{3-6}
    &  & 3 & NLI & 180 & 2 \\ \cline{3-6} 
    &  & 4 & STD & 520 & 6 \\ \cline{3-6} 
    &  & 5 & STD & 1021 & 8 \\ \hline
    \multirow{5}{*}{2} & \multirow{5}{*}{N} & 1 & STD & 212 & 2 \\ \cline{3-6} 
    &  & 2 & STD & 419 & 4 \\ \cline{3-6} 
    &  & 3 & NLI & 306 & 3 \\ \cline{3-6} 
    &  & 4 & NLI & 729 & 5 \\ \cline{3-6} 
    &  & 5 & STD & 741 & 9 \\ \hline
    \multirow{5}{*}{3} & \multirow{5}{*}{N} & 1 & NLI & 165 & 0 \\ \cline{3-6} 
    &  & 2 & NLI & 265 & 2 \\ \cline{3-6} 
    &  & 3 & STD & 764 & 10 \\ \cline{3-6} 
    &  & 4 & STD & 1189 & 10 \\ \cline{3-6} 
    &  & 5 & STD & 812 & 7 \\ \hline
    \multirow{5}{*}{4} & \multirow{5}{*}{N} & 1 & STD & 315 & 5 \\ \cline{3-6} 
    &  & 2 & NLI & 197 & 2 \\ \cline{3-6} 
    &  & 3 & STD & 610 & 6 \\ \cline{3-6} 
    &  & 4 & NLI & 576 & 3 \\ \cline{3-6} 
    &  & 5 & NLI & 382 & 3 \\ \hline
    \multirow{5}{*}{5} & \multirow{5}{*}{N} & 1 & NLI & 190 & 0 \\ \cline{3-6} 
    &  & 2 & STD & 598 & 8 \\ \cline{3-6} 
    &  & 3 & NLI & 247 & 3 \\ \cline{3-6} 
    &  & 4 & STD & 1186 & 5 \\ \cline{3-6} 
    &  & 5 & NLI & 431 & 5 \\ \hline
    \multirow{5}{*}{6} & \multirow{5}{*}{E} & 1 & NLI & 90 & 0 \\ \cline{3-6} 
    &  & 2 & STD & 197 & 2 \\ \cline{3-6} 
    &  & 3 & NLI & 91 & 0 \\ \cline{3-6} 
    &  & 4 & NLI & 410 & 1 \\ \cline{3-6} 
    &  & 5 & STD & 547 & 1 \\ \hline
    \multirow{5}{*}{7} & \multirow{5}{*}{E} & 1 & STD & 122 & 1 \\ \cline{3-6} 
    &  & 2 & NLI & 109 & 2 \\  \cline{3-6} 
    &  & 3 & STD & 169 & 1 \\  \cline{3-6} 
    &  & 4 & STD & 623 & 4 \\  \cline{3-6} 
    &  & 5 & NLI & 315 & 0 \\  \hline
    \multirow{5}{*}{8} & \multirow{5}{*}{E} & 1 & STD & 176 & 1 \\ \cline{3-6} 
    &  & 2 & NLI & 138 & 0 \\  \cline{3-6} 
    &  & 3 & STD & 201 & 1 \\  \cline{3-6} 
    &  & 4 & NLI & 484 & 5 \\  \cline{3-6} 
    &  & 5 & NLI & 206 & 0 \\  \hline
    \end{tabular}
    \label{tab-user}
\end{table}

\begin{figure}[htb]
\centering
\includegraphics[width=0.44\textwidth]{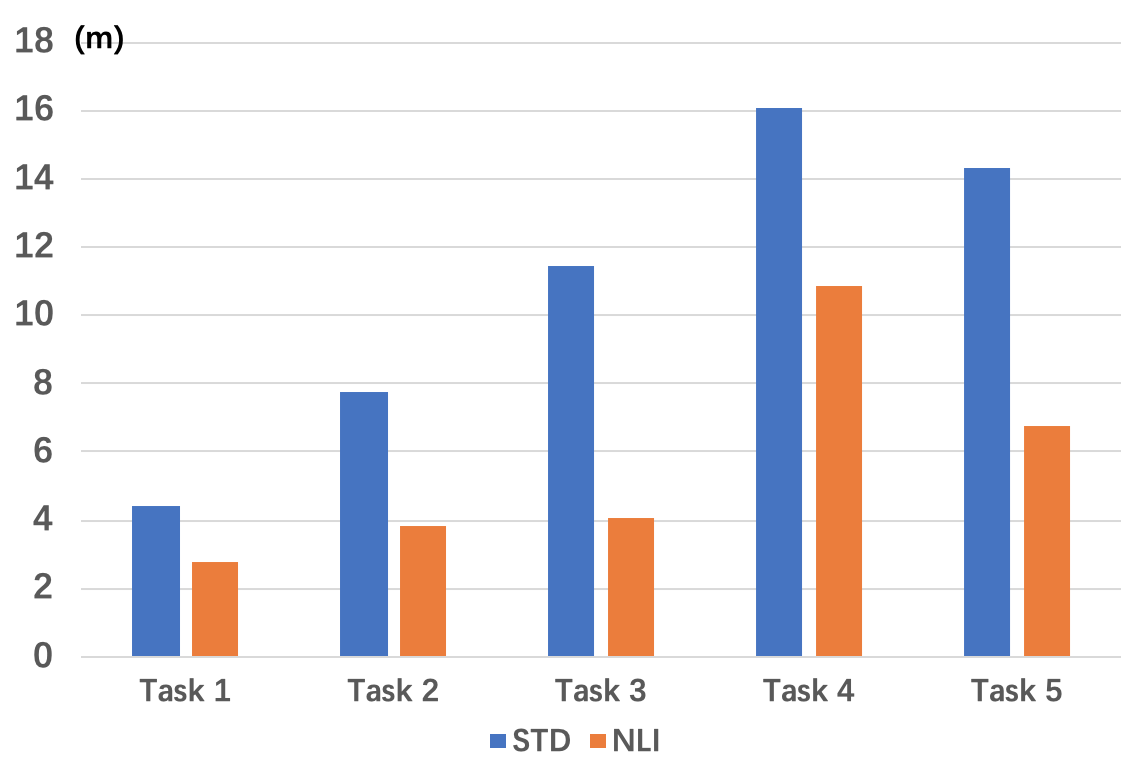}
\caption{Comparison between the average time newcomers spent in two coding environments\label{fig-compare1}}
\end{figure}

\begin{figure}[htb]
\centering
\includegraphics[width=0.44\textwidth]{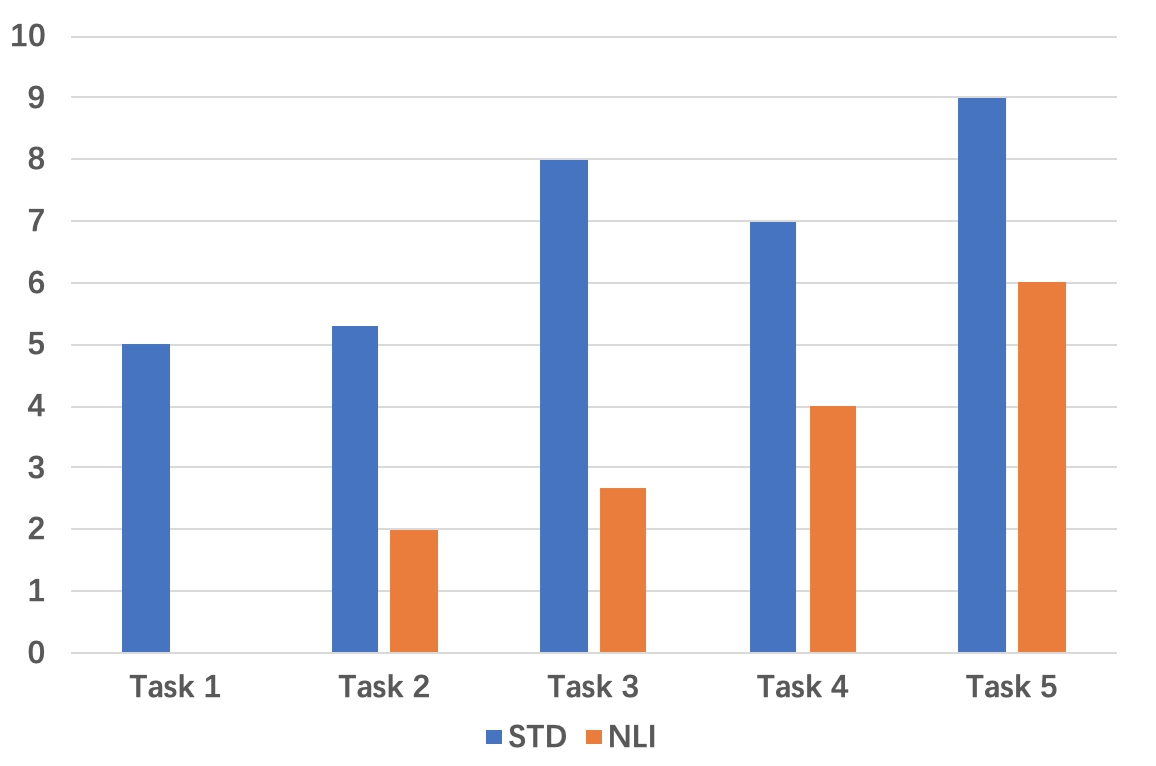}
\caption{Comparison between the average number of web pages newcomers visited in two coding environments} 
\label{fig-compare2}
\end{figure}

Figures \ref{fig-compare1} and \ref{fig-compare2} summarize the data from Table \ref{tab-user} for newcomers. 
It compares the average time (minutes) used and the number of web pages opened by the newcomers between two coding environments.
On average, newcomers without \textsc{NLI4j} spent 674 seconds and visited 6.5 web pages for each task, which is significantly larger than the number for participants using the plugin (317.7 seconds and 2.3 pages).
Notice that when newcomers using \textsc{NLI4j} met the first task (\emph{create blank workbook}), all of them solved the task without referring to any web sources.

However, such a difference is not that obvious in the expert group.
On average, an expert with \textsc{NLI4j} solved a task in 230.3 seconds, an expert without \textsc{NLI4j} solved a task in 290.7 seconds.
From later communication with the three experts, we found they were familiar with how to read and search library documentation, which can explain the number of web pages they opened was much smaller than the newcomer group.
However, all of the three experts confirmed the plugin is convenient when they forgot how to use a certain API.
Many participants reported that when they used the plugin, they could accomplish most tasks without external information and the visited web pages were only to confirm the solution.

For participants who solved tasks with \textsc{NLI4j}, we also asked them to record the rankings of the expressions they chose to complete the code patterns.
We wanted to know whether our synthesizer recommended useful expressions to the participants.
Table \ref{tab-interaction} shows the result, the second column lists the number of interactions for each task.
An interaction means there is a missing variable for users to provide or select from the list of recommended expressions.
We use two metrics, \textit{i.e.} MRR (Mean Reciprocal Rank) and Hit@1 to evaluate the quality of the recommendation.
Our synthesizer could not recommend useful expressions if the missing part must be specified by users.
For example, the only interaction in the first task is the name of the workbook, which could be any valid string.
\textsc{NLI4j} failed to recommend the desired string and get 0 for both MRR and Hit@1 metrics.
Actually, among all 13 interactions for the tasks, such conditions (arbitrary values of built-in types) happened 5 times.
For all the other interactions, \textsc{NLI4j} successfully recommended the desired expressions at a top-2 position.
In the third task, all four desired variables were recommended as the first choice.
On average, each task requires 2.6 interactions and the average MRR value is 0.54 and the value for Hit@1 is 0.46.

\begin{table}
\centering
\caption{Recommendation performance of the synthesizer\label{tab-interaction}}
\begin{tabular}{c c c c}
\hline
\bfseries Task ID & \bfseries \#Interactions & \bfseries MRR & \bfseries Hit@1\\
\hline
1 & 1 & 0 & 0\\
2 & 2 & 0.25 & 0\\
3 & 4 & 1.00 & 1.00\\
4 & 4 & 0.375 & 0.25\\
5 & 2 & 0.50 & 0.50\\
\hline
\bfseries Average & \bfseries 2.6 & \bfseries 0.54 & \bfseries 0.46\\
\hline
\end{tabular}
\end{table}

\subsubsection{Discussion}
All participants were asked to fill a simple survey after the controlled experiment.
The survey form is available in our published artifacts.
Form the survey result, we can see all participants agree that using \textsc{NLI4j} could improve their coding efficiency.
When asked to compare the input form of functional features with free-form natural language, most participants (6 out of 8) reported that they preferred functional features.
However, some participants raised the concern that for those functions not included in functional features, they could only turn to the free-form queries.
Besides, one of our participants mentioned that although free-form queries are flexible, however, composing such queries from scratch could be difficult for a newcomer.
He mentioned that some hint like auto-completion or our functional features would be very helpful when users described their requirements.

We also asked the participants to compare code patterns used in \textsc{NLI4j} with concrete code examples.
Overall, most of the participants (5 out of 8) preferred code patterns with two main reasons. 
First, code patterns gave a more clear hint for where to modify.
Second, participants believed that code patterns had higher quality and were more reliable since they were mined from multiple concrete examples.

\mybox{\emph{Answer for RQ4:}
The result of the controlled experiment shows that \textsc{NLI4j} can save half of the coding time for newcomers of a library.
For experienced developers, \textsc{NLI4j} can play the role of a prompter when they forget the usage of certain APIs.
Given a programming context, the recommended expressions from our synthesizer can effectively help developers fill the missing parts.
}
\subsection{Threats to Validity}
\textbf{Internal validity: }
Our four research questions covered the key steps in constructing NLI (\textit{i.e.}, functional feature extraction, code pattern mining, and the synthesizer).
However, we could not evaluate all the details in the implementation because our framework has a quite long workflow.
For example, we did not discuss parameter tuning for our frequent pattern mining algorithm.
In our current implementation, we set the frequency threshold as 5\% to mine code patterns and it works well on our datasets.
However, the best threshold may vary under different datasets.

For the case study, although we have considered the help of \textsc{NLI4j} varies for different users.
The total number of participants is relatively small. We plan to put our tool in the daily development of developers and collect more user data in our future work.

\textbf{External validity: }
We selected five libraries from different domains, which covered the front-end parsing tool, the back-end database, and popular toolkits.
The evaluation shows that our tool can mine accurate functional features (accuracy of 86.2\%) and high-quality code patterns.
However, since our tool is feature-oriented, its performance on libraries with clear features are usually better than the libraries which are designed as frameworks.
Furthermore, API invocation is not the only way of library reuse. 
Some libraries heavily rely on other design or syntax, such as design patterns and annotations (\textit{e.g.}, the OGM mechanism in \textit{neo4j}).
Thus, the first external validity is the generalization of our framework to other libraries.

We carefully chose the datasets in our experiment so the findings could be generalized as much as possible.
We selected Stack Overflow to extract functional features because it is one of the most popular platforms to search for programming tasks \citep{uclsurvey}.
There are a lot of discussions about API usage from the site, and many previous studies encourage us to select it as the corpus (\textit{e.g.}, \citep{treude16, examplestack}). 
Nonetheless, not all the libraries are active on Stack Overflow.
Although most of our design is not specific to Stack Overflow, the performance may differ when other forms of user discussions are used as input.
Regarding the codebase for mining code patterns, we downloaded all repositories with at least 5 stars from Github and the number of source code files is more than 105K.
In our experiment, we found a code corpus containing one hundred files is good enough to mine high-quality patterns.
However, we only evaluated on Java APIs and it may not be representative to all the languages and libraries.

\section{Related Work}
\label{related}
The idea of \textsc{NLI2Code} contributes to the large body of work on API comprehension and software reuse.
In addition, each of the three components has benefited from related work in the corresponding domain, which will be summarized in this section separately.

\subsection{Information extraction from software artifacts}
Several researchers have succeeded in extracting high-quality software specifications from software artifacts using NLP techniques.
\cite{zhong09} proposed an approach for inferring specifications from API documentation by detecting actions and resources through machine learning.
Their evaluation showed relatively high precision, recall, and F-scores for five software libraries, and indicated potential uses in bug detection.
\cite{concept} presented an NLP-based approach to extract and organize concepts from software identifiers in a WordNet-like structure through tokenization, part-of-speech tagging, dependency sorting, and lexical expansion.
\cite{api-tutorial} introduced an approach to select relevant tutorial fragments for APIs, which combined the topic model and the PageRank algorithm.
More closely to our goal, summarizing software artifacts with functional explanations, \cite{faq} designed an approach to extract FAQs from mailing lists and forums.
The approach applied the LDA algorithm to extract topic models from the data which are used for the creation of topic-specific FAQs.
\cite{task1} defined the concept of task as a specific programming action that has been described in the documentation.
Indexing long documents with high-level tasks can help users quickly locate the part they care about.
Furthermore, \cite{task2code} developed a tool that can map tasks to code snippets from Stack Overflow answers.
In \textsc{NLI2Code}, we normalize the free-form tasks into a set of pre-defined functional features and enhance concrete code examples into abstract code patterns, considering the quality of Stack Overflow code examples are controversial \citep{api-misuse}. 

\subsection{Code pattern mining}
Code patterns are abstract code examples with metavariables or other components to be completed by users.
Modern IDEs usually integrate relevant features to define widely-used code patterns, such as live template feature in IntelliJ IDEA and SnipMatch in Eclipse.
Several studies \citep{api2, api3} applied statistical methods to automatically mine code patterns since source code was shown to be highly repetitive \citep{naturalness}.
The common workflow for code pattern mining first abstracts source code into a well-designed data structure and then apply the corresponding frequent pattern mining algorithm.
\cite{fse14:idiom} presented \textsc{Haggis}, a system for mining code patterns that was built on techniques from statistical natural language processing.
\textsc{Haggis} transformed source code into abstract syntax trees and applied Bayesian probabilistic tree substitution technique to get code patterns.
The mined patterns were proved to be accurate and meaningful and the author mentioned part of the patterns were accepted by the Eclipse SnipMatch project.
To detect API misuse in online forums, \cite{api-misuse} developed a tool \textsc{ExampleCheck} to compare API usages in the forum with code patterns mined from large codebases.
The authors designed a data structure called the structured call sequence, which enriched API invocations with syntax like guard conditions and control flow statements.
Such enrichment is vital because most API misuses in online code examples suffer from missing guard conditions and exception handling and \textsc{ExampleCheck} could effectively find these misuses.

Compared with existing works, which are designed to solve a particular problem, \textsc{NLI2Code} is designed as an abstract framework, which does not specify the approach to mine code patterns.
Code abstraction designed in existing tools may rely on properties of their problems and cannot be easily generalized to others.

\subsection{Program synthesis from natural language}
Program synthesis is the task of automatically finding a program in the underlying programming language that satisfies the user intent expressed in the form of some specification \citep{synthesis-overview}.
This problem has been considered the holy grail of computer science since the inceptions of AI in the 1950s.
Program synthesis works diverse in the form of specification, including partial data structures \citep{pbp}, test cases \citep{pbe}, natural language \citep{t2api, keyword} and their combination \citep{IJCAI15}.
Despite its ambiguity, the natural language specification is the most flexible one and requires the smallest effort to compose.
Existing synthesis tools with natural language input either recommend related APIs \citep{demomatch, ASE17:bot} or compilable snippets \citep{usage}.
\cite{OOPSLA15} defined a free-form specification that allowed users to write natural language queries and use names of local variables.
Given a specification, they mapped it to a method and expanded the method with a PCFG model trained from large codebases.
\cite{codehint} developed a dynamic and interactive program synthesis tool \textsc{CodeHint}, which was integrated into the Eclipse IDE.
\textsc{CodeHint} allowed users to execute the recommended code snippets and refine the snippets iteratively.

\section{Conclusion}
\label{conclusion}
This paper promotes the concept of NLI (Natural Language Interface) for library reuse.
To construct and use NLI, we design a framework with three components (\textit{i.e.}, functional feature extractor, code pattern miner, and synthesizer).
We instantiate the three components as a tool \textsc{NLI4j} to reuse Java libraries.
The accuracy of our extracted functional features is 86.2\% and can cover 86.3\% of functionalities provided by the official tutorials.
By comparing with existing code pattern miner, \textsc{NLI4j} can mine more accurate and complete code patterns.
Finally, a controlled experiment with eight participants on five real-world tasks shows that our tool can save half of the coding time for newcomers of the library.
From the practical perspective, our framework promotes the efficiency of reusing libraries.
From the academic perspective, our framework lays out a design space of building the natural language interface for libraries, which would hopefully inspire research in this area.

\section*{Acknowledgment}
This paper is supported by National Natural Science Fund for Distinguished Young Scholars (No. 61525201) and General Program of National Natural Science Foundation of China (61972006).

\bibliographystyle{model5-names}
\bibliography{main.bib}

~\\

\noindent \textbf{Qi Shen} is a Ph.D candidate at the Software Institute, Peking University, under the supervision of Prof. Bing Xie and Associate Prof. Yanzhen Zou.
His research interests include software reuse and program synthesis.
He got a bachelor degree in computer science from Peking University in June, 2016. 
\\

\noindent \textbf{Shijun Wu} is a Ph.D. candidate at the Software Institute, Peking University.
His main research interests are software engineering and human computer interaction.
He got a bachelor degree in computer science from Peking University in June, 2019.  \\

\noindent \textbf{Yanzhen Zou} received her Ph.D. degree from Peking University in 2010.
She is currently an associate professor in in the School of Electronics Engineering and Computer Science, Peking University.
Her research interests focus on software engineering and information retrieval.
\\

\noindent \textbf{Zixiao Zhu} is currently a research scientist at IBM Research - China lab.
He received the B.Eng. degree in software engineering from East China Normal University, Shanghai, China, in 2011, and the Ph.D. degree in computer software and theory from Peking University, Beijing, China, in 2018.
His research interests include unstructured data mining and analytics, AIOps, log analytics, knowledge graph, and software reuse.
\\

\noindent \textbf{Bing Xie} is a professor in the School of Electronics Engineering and Computer Science, Peking University.
He is leading the Software Institute at Peking University.
He received his Ph.D. degree from the National University of Defense Technology.
He has more than 80 publications at major conferences in software engineering including FSE and POPL.
His research interests include software reuse and formatting system.
\\

\end{document}